\documentclass[aps,nofootinbib,prd]{revtex4-2}

\usepackage{epsfig,epsf}
\usepackage{amsmath}
\usepackage{amsthm}
\usepackage{amsfonts}
\usepackage{amssymb}
\usepackage{dsfont}
\usepackage{multirow}
\usepackage{appendix}
\usepackage{slashed}
\usepackage[active]{srcltx}
\usepackage{psfrag}
\usepackage{subfigure}
\usepackage[colorlinks,citecolor=blue,urlcolor=red,linkcolor=red]{hyperref}

\begin{document}
\title{{\Large{\bf Holographic study of the scalar meson $K_0^*(1430)$ with dynamical longitudinal confinement }}}

\author{\small
O. Hemmati \footnote {e-mail: hemmati.osra@ph.iut.ac.ir }, R.
Khosravi \footnote {e-mail: rezakhosravi@shirazu.ac.ir }}

\affiliation{{\small $^a$ Department of Physics, Isfahan University
of Technology, Isfahan 84156-83111, Iran   }\\
{\small $^b$ Department of Physics, School of Science, Shiraz
University, Shiraz 71946-84795, Iran} }

\begin{abstract}
Within the framework of light-front holographic QCD, where
transverse confinement is inherently realized through holography,
the structure and mass spectrum of the light scalar meson
$K_0^{*}(1430)$ are investigated by employing the invariant mass
ansatz and enhancing the model with a longitudinal confinement
potential (specifically, the 't Hooft equation and the analytically
solvable LMZV potential). The combined transverse and longitudinal
confinement dynamics lead to more reliable calculations of key
observables, including decay constants, parton distribution
function, $\xi$-moments, and twist-$2$ and twist-$3$ distribution
amplitudes. Transition form factors for the semileptonic decay $B
\rightarrow K_0^{*} \ell^+ \ell^-$ are computed from these
amplitudes, yielding predictions for differential branching ratios
and longitudinal lepton polarization asymmetries as functions of
$q^2$. The results are compared with other theoretical models,
offering valuable insights into the non-perturbative dynamics of
QCD. The predicted observables provide theoretically motivated
benchmarks for future experimental investigations of $B\rightarrow
K_0^*(1430)\ell^+\ell^-$ decays at LHCb and Belle II.
\end{abstract}


\maketitle

\section{Introduction}
Understanding the internal structure of light scalar mesons is
crucial for explaining nonperturbative phenomena in QCD,
specifically confinement and dynamical chiral-symmetry breaking
\cite{Narison}. Scalar mesons with vacuum quantum numbers ($J^{\rm
PC} = 0^{++}$) present distinctive theoretical and experimental
challenges, as their microscopic structure remains a subject of
active debate. Experimentally, the analysis is complicated by
several factors: (i) large decay widths leading to significant
resonance overlap; (ii) the simultaneous opening of multiple decay
channels within narrow mass intervals; (iii) pronounced cusp
structures generated by threshold effects from $K\bar{K}$ and
$\eta\eta$; and (iv) the expected contributions of non-$q\bar{q}$
configurations, such as glueballs and multiquark states, in the mass
region below $1800~\text{MeV}$ \cite{SpToAm}. On the theoretical
side, a variety of interpretations have been proposed, ranging from
conventional $q\bar{q}$ states to exotic configurations such as
tetraquarks \cite{RLJa}, hadronic molecules \cite{UGMe}, hybrids
\cite{DHoJMa}, and glueballs \cite{HFrMGe}. Since these phenomena
arise in the strongly coupled regime of QCD, a quantitatively
controlled and systematically improvable framework remains an
essential objective.Although no data are yet available

Among strange scalar mesons, $K_0^*(1430)$ occupies a particularly
important position. High-statistics measurements in $pp$
annihilation, $\pi N$ scattering, central production, $J/\Psi(1S)$
decays, and $\gamma\bar{\gamma}$ formation \cite{SpToAm} have firmly
established this resonance with mass and width
$(1425\pm50)\,\text{MeV}$ and $(270\pm80)\,\text{MeV}$, respectively
\cite{pdg1}. It is commonly assigned to the lowest-lying strange
scalar $q\bar q$ nonet with $L=S=1$. Yet a quantitative
understanding of its internal structure requires a theoretical
approach in which confinement and chiral symmetry breaking emerge in
a mutually consistent manner.

Light-front holographic QCD (LFHQCD), motivated by the AdS/CFT
correspondence \cite{BroTer}, offers a semiclassical approximation
to QCD in which confinement is encoded in an effective light-front
Schr\"odinger equation governing transverse dynamics. In this
approach, a correspondence between the fifth dimension of AdS space
and the transverse separation of the quark-antiquark pair leads to
realistic hadron spectra and wave functions
\cite{HonYooSt,Rady,HRGAVR,GriRad}. However, while the transverse
structure is tightly constrained, the longitudinal component of the
light-front wave function is not uniquely determined and requires
additional dynamical input.

A commonly used prescription for incorporating quark masses is the
invariant mass ansatz (IMA) \cite{BrTe1}, in which the transverse
kinetic term is modified by adding quark-mass contributions.
Although phenomenologically successful in many applications, this
approach has been shown to be inconsistent with the
Gell-Mann--Oakes--Renner (GMOR) relation \cite{VarLi1,GMOR},
demonstrating that, in the case of the pion, longitudinal dynamics
cannot be treated purely as a kinematical correction. This
observation has motivated the development of dynamical approaches to
longitudinal confinement, which provide a more complete theoretical
framework for describing the longitudinal structure of light-front
wave functions (LFWF).

The 't Hooft equation, derived in large-$N_c$ QCD within $(1+1)$
dimensions \cite{tHooft}, provides a longitudinal bound-state
equation that incorporates chiral-symmetry-breaking effects at the
Hamiltonian level. When combined with the holographic transverse
equation, it yields a successful description of hadron spectroscopy
\cite{AhmMacKKa}. However, the nonlocal structure of the 't Hooft
interaction precludes analytical solutions. To overcome this
limitation, an analytically solvable longitudinal potential (LMZV)
\cite{LiMaZhVa} has been proposed, it preserves the correct endpoint
behavior of parton distributions and reproduces the GMOR relation in
the vicinity of the chiral limit. Compared with other longitudinal
confinement models \cite{ChabHill,Glazek,VarLi2}, it offers a
tractable and systematically improvable framework for incorporating
longitudinal confinement dynamics.

In this work, we determine the mass spectrum and LFWF of the
$K_0^*(1430)$ meson within the framework of LFHQCD, explicitly
incorporating its helicity-dependent spin structure. The IMA is
first employed as a baseline description of quark-mass effects and
is subsequently supplemented by the dynamical 't Hooft and LMZV
longitudinal interactions. This framework enables a direct and
controlled comparison between the phenomenological invariant-mass
ansatz and the dynamical longitudinal confinement models under
identical input parameters. Our objective is not to argue that
dynamical longitudinal confinement is indispensable for describing
the scalar meson $K_0^*(1430)$, but rather to investigate how
different implementations of longitudinal dynamics influence the
nonperturbative structure of the meson and the corresponding
predictions for physical observables. In particular, we identify
which observables are sensitive to the underlying longitudinal
dynamics and which remain comparatively robust against different
modeling prescriptions. Using the resulting LFWFs, we compute the
decay constants, parton distribution function (PDF), $\xi$-moments,
and twist-$2$ and twist-$3$ distribution amplitudes (DAs) of the
$K_0^*(1430)$ meson.

Heavy-meson decays provide an additional and complementary probe of
scalar-meson structure. Experimental studies of $B$ decays into
scalar states have been reported by the Belle \cite{Garm,Belle},
BaBar \cite{Aube,BaBar}, and LHCb \cite{LHCb} collaborations.
Although no data are yet available for the semileptonic channel $B
\to K_0^*(1430)\,\ell^+\ell^-$, theoretical predictions for its
transition form factors have been obtained using the light-cone sum
rule (LCSR)
\cite{Khosravi1,Khosravi2,ZGWan,ZGWang,HanWuFu,SunLi,WanAsl}, the
QCD sum rules (QCDSR) \cite{NGhahramany,AliAzi,MZYan}, the
perturbative QCD (PQCD) approach \cite{RunHui}, the light-front
quark model (LFQM) \cite{ChGeLi}, the covariant light-front (CLF)
approach \cite{CheChuHwa}, and the minimal supersymmetric standard
model (MSSM) \cite{AsluWang}. In the present work, we compute the
corresponding transition form factors, differential branching
ratios, and longitudinal lepton polarization asymmetries for $B \to
K_0^*(1430)\,\ell^+\ell^-$ with $\ell=\mu,\tau$, utilizing the
holographic twist-$2$ and twist-$3$ DAs derived in this study.

Although no dedicated experimental measurements of the rare decay
$B\rightarrow K_0^*(1430)\ell^+\ell^-$ have been reported to date,
its experimental investigation is well motivated within the ongoing
flavor-physics programs of both the LHCb and Belle~II
experiments~\cite{LHCb2021,LHCb2025,Belle2020}. Rare semileptonic
$b\rightarrow s\ell^+\ell^-$ transitions constitute one of the
primary objectives of these facilities~\cite{LHCb2012}, while scalar
$K\pi$ contributions are routinely incorporated in amplitude
analyses of exclusive $B$ decays~\cite{LHCb2019}. Consequently, the
theoretical predictions presented in this work should be viewed as
benchmarks for future experimental studies rather than quantities
that can presently be confronted with data.

This paper is organized as follows: Sec. II constructs the
scalar-meson LFWF, detailing the helicity-dependent spin structure
and the dynamical quark-antiquark component within LFHQCD. In Sec.
III, we investigate dynamical realizations of the longitudinal mode
by comparing the phenomenological IMA with the nonperturbative 't
Hooft and LMZV longitudinal interactions; the latter is selected for
its analytic solvability and consistency with the GMOR relation.
Subsequently, Sec. IV derives the twist-$2$ and twist-$3$ DAs of the
$K_0^*(1430)$ meson, analyzing the sensitivity of DA shapes to the
underlying longitudinal dynamics. Finally, Sec. V provides a
comprehensive qualitative and numerical analysis based on three
distinct implementations of the LFWF within the IMA, 't Hooft, and
LMZV frameworks, each realizing longitudinal confinement and chiral
symmetry breaking through different mechanisms. It also includes the
determination of decay constants and $\xi$-moments, along with a
detailed study of the PDF and holographic DAs. Based on the
calculated DAs, we present final predictions for the transition form
factors, differential branching ratios, and longitudinal
polarization asymmetries in the rare semileptonic decay $B \to
K_0^*(1430)\, \ell^+ \ell^-$ as functions of the momentum-transfer
squared $q^2$.

\section{Light-Front Wave Function for Scalar Mesons}
The LFWF for a meson in momentum space is defined as:
\begin{equation}\label{eq201}
\Psi_{h,\bar{h}}(\alpha, \mathbf{k}_{\perp}) = N \,
S_{h,\bar{h}}(\alpha, \mathbf{k}_{\perp}) \, \psi(\alpha,
\mathbf{k}_{\perp}),
\end{equation}
where $\alpha$ denotes the longitudinal momentum fraction carried by
the quark, while $\mathbf{k}_{\perp}$ represents the transverse
momentum. The symbols $h$ and $\bar{h}$ correspond to the helicity
states of the quark and antiquark, respectively. In this expression,
$S_{h,\bar{h}}$ is the spin wave function, and $\psi$ describes the
dynamical properties of the quark-antiquark pair. Together,
$S_{h,\bar{h}}$ and $\psi$ determine the helicity structure and
dynamical characteristics of the meson \cite{LepBro,DosGou}. The
normalization factor $N$ ensures that the total probability is
unity:
\begin{equation}\label{eq202}
\sum_{h,\bar{h}} \int_0^1 d\alpha \int \frac{d^2
\mathbf{k}_{\perp}}{16\pi^3} \left|\Psi_{h,\bar{h}}(\alpha,
\mathbf{k}_{\perp})\right|^2 = 1.
\end{equation}

For a scalar meson with mass $m$ and longitudinal momentum $p^{+}$,
the spin wave function $S_{h,\bar{h}}$, constrained by parity and
quark content, is defined as \cite{JCotanch}:
\begin{equation}\label{eq203}
S_{h,\bar{h}}(\alpha, \mathbf{k}_{\perp}) = \frac{\bar{u}_{h}(\alpha
p^+, \mathbf{k}_{\perp})}{\sqrt{\alpha}} \,(\not\!p + m)
\,\frac{v_{\bar{h}}(\bar{\alpha} p^+,
-\mathbf{k}_{\perp})}{\sqrt{\bar{\alpha}}}\,,
\end{equation}
where $\bar{\alpha}=1-\alpha$ is the momentum fraction carried by
the antiquark. Here, $u_{h}$ and $v_{\bar{h}}$ are the light-front
helicity spinors for the quark and antiquark, respectively. The
explicit forms of these spinors for positive and negative helicities
are given by \cite{LepBro}:
\begin{equation}\label{eq204}
u_{\uparrow,\downarrow} = \frac{1}{\sqrt{p^{+}_{q}}}\left( p^{+}_{q}
+ m_{q}\gamma_0 + {\alpha}_{\perp}\cdot \mathbf{k}_{\perp}
\right)\chi_{\uparrow,\downarrow}, \qquad v_{\uparrow,\downarrow} =
\frac{1}{\sqrt{p^{+}_{\bar{q}}}}\left( p^{+}_{\bar{q}}- m_{\bar
q}\gamma_0 + {\alpha}_{\perp}\cdot \mathbf{k}_{\perp}
\right)\chi_{\downarrow,\uparrow},
\end{equation}
where $p^{+}_{q}$ ($p^{+}_{\bar{q}}$) denotes the longitudinal
momentum of the quark $q$ (antiquark $\bar{q}$), with masses $m_{q}$
and $m_{\bar{q}}$, respectively. The operator ${\alpha}_{\perp}\cdot
\mathbf{k}_{\perp} = \gamma_0(\gamma_1 k_1+\gamma_2 k_2)$, and
$\chi_{\uparrow,\downarrow}$ are the eigenstates of the projection
operator:
\begin{equation}\label{eq205}
\chi_{\uparrow} = \frac{1}{\sqrt{2}}\begin{pmatrix} 1 \\ 0 \\ 1 \\ 0
\end{pmatrix}, \qquad \chi_{\downarrow} =
\frac{1}{\sqrt{2}}\begin{pmatrix} 0 \\ 1 \\ 0 \\ -1 \end{pmatrix}.
\end{equation}

By substituting the light-front helicity spinors from Eq.
(\ref{eq204}) into Eq. (\ref{eq203}), and evaluating the resulting
Dirac structure for a scalar configuration, one obtains the explicit
form of the spin wave function $S_{h,\bar{h}}$ for the $K_{0}^{*}$
meson:
\begin{equation}\label{eq206}
S_{h,\bar{h}}(\alpha, \mathbf{k}_{\perp}) = \frac{1}{\alpha
\bar{\alpha}}\left\{ \mathcal{M}_{\pm}\,k_{\perp} e^{\mp i\theta}
\delta_{h, \pm}\delta_{\bar{h}, \pm} + \left(\mathcal{M}^{2} -
k_{\perp}^{2} +
m_{K_{0}^{*}}\,\bar{\mathcal{M}}_{-}\right)\delta_{h, \pm}
\delta_{\bar{h},\mp}\right\},
\end{equation}
where $\mathcal{M}_{\pm}=\pm m_{s} \mp m_{\bar{q}} \pm
m_{K_{0}^{*}}$, $\mathcal{M}^{2}=\alpha \bar{\alpha}\,
m^2_{K_{0}^{*}}-m_{s} m_{\bar{q}}$,
$\bar{\mathcal{M}}_{\pm}=\bar{\alpha} m_s \pm\alpha m_{\bar{q}}$,
and $k_{\perp} e^{\pm i\theta}$ denotes the complex representation
of the transverse momentum vector $\mathbf{k}_{\perp}$.

Based on a relativistic semiclassical first approximation to QCD, an
effective Schr\"odinger-like equation in $(3+1)$ dimensions governs
the wave function $\psi$, which describes quark-antiquark bound
states in transverse momentum space \cite{TerBrod}:
\begin{equation}\label{eq207}
\left[\frac{m_q^2}{\alpha}+\frac{m_{\bar{q}}^2}{\bar{\alpha}}+\frac{k_{\perp}^2}{\alpha
\bar{\alpha}} +U \right] \psi = m^2 \psi,
\end{equation}
where $U$ is the effective light-front confinement potential.
Performing a transverse Fourier transform to relative coordinate
$\mathbf{r}_{\perp}$-space yields:
\begin{equation}\label{eq208}
\left[\frac{m_q^2}{\alpha}+\frac{m_{\bar{q}}^2}{\bar{\alpha}}-\frac{1}{\alpha
\bar{\alpha}}\nabla^{2}_{\perp} +U \right] \psi(r, \varphi, \alpha)
= m^2 \psi(r, \varphi, \alpha),
\end{equation}
where $\nabla^{2}_{\perp}=\frac{\partial^2}{\partial
r^2}+\frac{1}{r}\frac{\partial}{\partial
r}+\frac{1}{r^2}\frac{\partial^2}{\partial \varphi^2}$ is the
transverse Laplacian. Here, $r$ denotes the transverse separation
between the quark and antiquark, and $\varphi$ is the polar angle in
$\mathbf{r}_{\perp}$-space, with $\mathbf{r}_{\perp}=r
e^{i\varphi}$.

Introducing the boost-invariant variable $\zeta = \sqrt{\alpha
\bar{\alpha}}\, r$ allows the quark-antiquark wave function to
factorize into transverse and longitudinal modes:
\begin{equation}\label{eq209}
\psi(\zeta, \varphi, \alpha) = \frac{\phi(\zeta)}{\sqrt{2 \pi
\zeta}}\, e^{i L \varphi}\, X(\alpha),
\end{equation}
where $L = |L^{\rm{max}}_z |$ is the orbital angular momentum
quantum number. Substituting Eq. (\ref{eq209}) into Eq.
(\ref{eq208}) leads to:
\begin{equation}\label{eq210}
\left[\frac{m_q^2}{\alpha}+\frac{m_{\bar{q}}^2}{\bar{\alpha}}-\frac{\partial^2}{\partial\zeta^2}
- \frac{1 - 4L^2}{4\zeta^2} + U \right] \phi(\zeta) X(\alpha) = m^2
\phi(\zeta) X(\alpha).
\end{equation}

In the chiral limit of LFHQCD, the quark and antiquark masses
vanish, $m_q=m_{\bar{q}}=0$, and the potential $U$ reduces to a
function of $\zeta$ alone, denoted by $U_{\perp}(\zeta)$. As a
result, only the transverse mode remains dynamical and is governed
by the holographic Schr\"odinger-like equation:
\begin{equation} \label{eq211}
\left[-\frac{d^2}{d\zeta^2} - \frac{1 - 4L^2}{4\zeta^2} +
U_{\perp}(\zeta) \right] \phi(\zeta) = m^2_{\perp} \phi(\zeta).
\end{equation}
According to LFHQCD \cite{TerSJBr}, identifying $\zeta$ with the
fifth dimension $z$ of $\rm{AdS}_5$ space maps this equation to the
wave equation for spin-$J$ modes propagating in $\rm{AdS}_5$,
modified by a quadratic dilaton field. The resulting light-front
holographic potential is fixed by the soft-wall condition
\cite{TeramBrod}:
\begin{equation} \label{eq212}
U_{\perp}^{\rm{LFH}}(\zeta) = \kappa^4 \zeta^2 + 2\kappa^2(J-1),
\end{equation}
with $J=L+S$. The universal transverse confinement scale $\kappa$
determines both the confinement strength and the meson mass spectrum
in the chiral limit. For this potential, the mass spectrum and
eigenfunctions are given by \cite{DosErl}:
\begin{align} \label{213}
m^2_{\perp}(n_{\perp}, L, S) &= 2\kappa^2 \left(2n_{\perp} + 2L + S \right), \\
\phi_{n_{\perp}}^{(L)}(\zeta) &=
\sqrt{\frac{2n_{\perp}!}{(n_{\perp}+L)!}} \kappa^{1+L} \zeta^{1/2+L}
\exp \left(-\frac{\kappa^2 \zeta^2}{2}\right)
L_{n_{\perp}}^{(L)}(\kappa^2 \zeta^2),
\end{align}
where $n_{\perp}=0,1,2,\dots$ is the radial quantum number, and
$L_{n_{\perp}}^{(L)}$ denotes the associated Laguerre polynomials.

It should be noted that the longitudinal mode
$X(\alpha)=\sqrt{\alpha\bar{\alpha}}\,\chi(\alpha)$ is not dynamical
within LFHQCD and is therefore not determined by Eq. (\ref{eq211}).
Instead, it is constrained through the holographic mapping of the EM
(or gravitational) pion form factor between physical space-time and
$\rm{AdS}_5$. This procedure uniquely leads to
$X(\alpha)=\sqrt{\alpha\bar{\alpha}}$ \cite{BrTe,BrTe02}, and hence:
\begin{equation} \label{eq214}
\chi^{\rm{LFH}}(\alpha)=1.
\end{equation}

Therefore, the quark-antiquark wave function in the holographic
light-front approach for a meson with $L=1$ takes the form:
\begin{equation} \label{eq215}
\psi^{\rm{LFH}}_{n_{\perp}}(\zeta, \varphi, \alpha)=
\sqrt{\frac{n_{\perp}!}{(n_{\perp}+1)!}}\frac{\kappa^{2} \zeta
e^{i\varphi}}{\sqrt{\pi}} \sqrt{\alpha \bar{\alpha}} \exp
\left(-\frac{\kappa^2 \zeta^2}{2}\right)\,
L_{n_{\perp}}^{(1)}(\kappa^2 \zeta^2).
\end{equation}

\section{Dynamical Realizations of the Longitudinal Mode}

In this section we investigate the longitudinal structure of the
scalar meson $K_0^*(1430)$ using three complementary approaches: the
IMA, the 't Hooft model, and the LMZV framework. The IMA
incorporates quark-mass effects through a phenomenological
modification of the invariant mass without introducing an explicit
longitudinal confining interaction. In contrast, the 't Hooft model
generates a nonlocal longitudinal confining kernel derived from
large-$N_c$ QCD in $(1+1)$ dimensions, providing a dynamical
realization of confinement and chiral symmetry breaking. Although
its solutions must be obtained numerically, it captures essential
nonperturbative features of longitudinal dynamics. To obtain an
analytically tractable formulation, the LMZV approach introduces a
solvable longitudinal potential that closely reproduces the behavior
of the 't Hooft model and remains consistent with the GMOR relation
near the chiral limit. Taken together, these approaches enable a
systematic construction of the full light-front wave function by
consistently combining transverse and longitudinal dynamics.

\subsection{Invariant Mass Ansatz (IMA)}
The IMA provides a widely used phenomenological prescription for
incorporating finite quark masses within light-front holography
without introducing an explicit longitudinal interaction. In this
approach the LFWF, obtained in the chiral limit, is modified through
the invariant mass of the quark-antiquark system.

We first Fourier transform the transverse wave function in Eq.
(\ref{eq215}) to $\mathbf{k}_\perp$-space for $n_\perp=0$, obtaining
\begin{equation}\label{eq301}
\psi^{\rm LFH}(\alpha, \mathbf{k}_\perp) =
\frac{\sqrt{4\pi}\, k_\perp\, e^{i\theta}} {\alpha \bar{\alpha}\,
\kappa^2} \exp\!\left(-\frac{k_\perp^2}{2 \alpha \bar{\alpha}
\kappa^2}\right).
\end{equation}

The central assumption of the IMA is to replace the massless
invariant mass by its massive counterpart \cite{BrTe1},
\begin{equation}\label{eq302}
\frac{k_\perp^{2}}{\alpha \bar{\alpha}}
\;\longrightarrow\; \frac{k_\perp^{2}}{\alpha \bar{\alpha}} +
\frac{m_q^2}{\alpha} + \frac{m_{\bar q}^2}{\bar{\alpha}} .
\end{equation}

Implementing this substitution in Eq. (\ref{eq301}) yields the
modified longitudinal factor
\begin{equation}\label{eq303}
\chi^{\rm IMA}(\alpha) = \exp\!\left( -\frac{m_q^2}{2\kappa^2\alpha}
-\frac{m_{\bar q}^2}{2\kappa^2\bar{\alpha}} \right),
\end{equation}
and consequently the full quark-antiquark wave function for a meson
with $L=1$ becomes
\begin{equation}\label{eq304}
\psi_{n_\perp}^{\rm IMA} = \psi^{\rm
LFH}_{n_\perp}(\zeta,\varphi,\alpha)\, \chi^{\rm IMA}(\alpha).
\end{equation}

This ansatz introduces an $\alpha$-dependent exponential suppression
that smoothly disappears in the chiral limit $(m_q,m_{\bar q}\to0)$,
thereby recovering the massless holographic wave function. Despite
its computational simplicity, the IMA does not dynamically implement
chiral symmetry breaking and may exhibit tension with chiral
constraints such as the GMOR relation.

\subsection{The 't Hooft Model}
A dynamical realization of the longitudinal interaction can be
obtained by modeling the confinement kernel using the large-$N_c$
limit of QCD in $(1+1)$ dimensions. Within this framework the
effective potential in the light-front Schr\"odinger equation
(\ref{eq207}) is assumed to be separable,
\begin{equation}\label{eq305}
U(\zeta, \alpha) = U_\perp(\zeta) +
U_\parallel(\alpha),
\end{equation}
where $U_\perp(\zeta)$ governs the transverse dynamics and
$U_\parallel(\alpha)$ encodes the longitudinal interaction. While
$U_\perp(\zeta)$ is determined by the holographic equation
(\ref{eq211}), the longitudinal mode satisfies
\begin{equation}\label{eq306}
\left[ \frac{m_q^2}{\alpha} + \frac{m_{\bar q}^2}{\bar{\alpha}} +
U_\parallel(\alpha) \right] \chi(\alpha) = m_\parallel^2\,
\chi(\alpha),
\end{equation}
together with the mass decomposition $m^2=m_\perp^2+m_\parallel^2$,
where $m$ denotes the total meson mass.

In the 't Hooft model \cite{tHooft}, the longitudinal kernel is
given by the principal-value integral
\begin{equation} \label{eq307}
U_\parallel(\alpha)\chi(\alpha) = \frac{g^2}{\pi}\, \mathcal{P} \int
d\alpha'\, \frac{\chi(\alpha)-\chi(\alpha')} {(\alpha-\alpha')^2},
\end{equation}
where $g$ denotes the longitudinal confinement scale and
$\mathcal{P}$ indicates the Cauchy principal value. This nonlocal
interaction generates confinement along the longitudinal direction
and dynamically implements chiral symmetry breaking in the
large-$N_c$ limit.

The resulting integral equation does not admit analytic solutions.
For numerical implementation, the longitudinal wave function is
expanded in an orthonormal basis $f_n(\alpha)$ \cite{ChabHill},
\begin{equation} \label{eq308}
f_n(\alpha) = \mathcal{N}_n\, \alpha^{\frac{\beta}{2}}\,
\bar{\alpha}^{\frac{\gamma}{2}} P_{n}^{(\gamma,\beta)}(2\alpha-1),
\end{equation}
with normalization factor
\begin{equation}\label{eq308b}
\mathcal{N}_n = \sqrt{2n + \gamma + \beta + 1}
\sqrt{\frac{\Gamma(n+1)\Gamma(n+\gamma+\beta+1)}
{\Gamma(n+\gamma+1)\Gamma(n+\beta+1)}}.
\end{equation}

Here $P_n^{(\gamma,\beta)}$ are Jacobi polynomials,
$\beta=2m_q/\sigma$, and $\gamma=2m_{\bar q}/\sigma$, where $\sigma$
characterizes the longitudinal scale.

The longitudinal eigenfunctions are expanded in the basis
$\{f_n(\alpha)\}$ as
\begin{equation}\label{eq309}
\chi^{\text{'t Hooft}}_{n_\parallel}(\alpha) = \sum_n
c_n(n_\parallel)\, f_n(\alpha),
\end{equation}
where the coefficients $c_n$ depend on the orbital quantum number
$n_\parallel$ and are obtained by diagonalizing the kernel of the 't
Hooft equation in the chosen basis. The expansion converges rapidly,
so that only a few basis states are required to achieve numerical
stability. The input parameters employed in the numerical
calculations, including the transverse confinement scale $\kappa$,
the longitudinal confinement parameters $g$ and $\sigma$, and the
quark masses $m_u$ and $m_s$, are summarized in Table \ref{T301}.
\begin{table}[th]
\caption{Parameter set used in the numerical analysis (in GeV),
including the transverse confinement scale $\kappa$, the
longitudinal confinement parameters $g$ and $\sigma$, and the quark
masses $m_u$ and $m_s$.} \label{T301}
\begin{ruledtabular}
\begin{tabular}{ccccc}
$\kappa$ & $g$ & $\sigma$ & $m_u$ & $m_s$ \\
\hline $0.524 \pm 0.023$ & $0.128$ & $0.120 \pm 0.012$ & $0.232 \pm
0.025$ & $0.357 \pm 0.036$
\end{tabular}
\end{ruledtabular}
\end{table}
The parameters $\kappa$, $g$, and $\sigma$, as well as the
strange-quark mass $m_s$, are adopted from Refs.
\cite{DosErl,AhmMacKKa}. In contrast, the effective light-quark mass
$m_u$ is calibrated in the present work by requiring that the
calculated mass of the scalar meson $K_0^*(1430)$ reproduces its
experimental value, while all the remaining model parameters are
kept fixed.

The resulting values of the first eight coefficients
$c_n(n_\parallel)$, together with the corresponding longitudinal
mass eigenvalues $m_\parallel(n_\parallel)$ for the $s\bar{u}$ kaon
family up to $n_\parallel=3$, are listed in Table~\ref{T302}.
\begin{table}[ht]
\caption{Expansion coefficients $c_n$ for the first eight basis
functions and the longitudinal mass eigenvalue
$m_{\parallel}(n_{\parallel})$.} \label{T302}
\begin{ruledtabular}
\begin{tabular}{cccccccccc}
$n_{\parallel}$ & $c_0$ & $c_1$ & $c_2$ & $c_3$ & $c_4$ & $c_5$ & $c_6$ & $c_7$ & $m_{\parallel}\,(\rm GeV)$ \\
\hline
$0$ & $0.991$  & $0.012$  & $-0.128$ & $-0.018$ & $0.029$ & $0.006$  & $0.002$  & $-0.000$ & $0.636$ \\
$1$ & $0.011$  & $-0.940$ & $-0.046$ & $0.321$  & $0.055$ & $-0.089$ & $-0.022$ & $0.011$  & $0.713$ \\
$2$ & $-0.127$ & $0.026$  & $-0.852$ & $-0.098$ & $0.460$ & $0.113$  & $-0.144$ & $-0.051$ & $0.768$ \\
$3$ & $-0.001$ & $0.313$  & $-0.079$ & $0.714$  & $0.117$ & $-0.564$ & $-0.098$ & $0.209$  & $0.820$ \\
\end{tabular}
\end{ruledtabular}
\end{table}

The complete mass spectrum follows from
\begin{equation}\label{eq310}
m^2(n_\perp, n_\parallel, L, S) = 2\kappa^2 (2n_\perp + 2L + S) +
m_\parallel^2(n_\parallel).
\end{equation}

The full quark-antiquark wave function in the 't Hooft framework is
therefore
\begin{equation}\label{eq311}
\psi^{\text{'t Hooft}}_{n_\perp, n_\parallel} = \psi^{\rm
LFH}_{n_\perp}(\zeta, \varphi, \alpha)\, \chi^{\text{'t
Hooft}}_{n_\parallel}(\alpha).
\end{equation}

This construction provides a fully dynamical description of meson
structure, combining holographic transverse confinement with
longitudinal dynamics derived from large-$N_c$ QCD.

\subsection{The LMZV Approach}
To overcome the absence of analytic solutions in the 't Hooft
equation, a solvable longitudinal confinement potential was
introduced in Ref. \cite{LiMaZhVa},
\begin{equation}\label{eq312}
U_\parallel(\alpha) = -\sigma^2 \frac{d}{d\alpha} \left(
\alpha\bar{\alpha} \frac{d}{d\alpha} \right),
\end{equation}
where $\sigma$ denotes the longitudinal confinement scale. Notably,
it reproduces the GMOR relation near the chiral limit, thereby
maintaining consistency with fundamental aspects of chiral symmetry
breaking \cite{VarLi1}.

Solving the longitudinal eigenvalue equation (\ref{eq306}) with the
potential above yields analytic expressions for both eigenvalues and
eigenfunctions \cite{VarLi,TerBrodsky},
\begin{align}\label{eq313}
m^2_{\parallel}(n_\parallel) &= \sigma (2n_\parallel +
1)(m_q + m_{\bar q}) + \sigma^2 n_\parallel (n_\parallel + 1) + (m_q
+ m_{\bar q})^2,
\\
\chi_{n_\parallel}^{\rm LMZV}(\alpha) &= \mathcal{N}_{n_\parallel}
\alpha^{\frac{\beta}{2}} \bar{\alpha}^{\frac{\gamma}{2}}
P_{n_\parallel}^{(\gamma,\beta)} (2\alpha - 1),
\end{align}
where the normalization factor $\mathcal{N}(n_{\parallel})$ is given
by:
\begin{equation}\label{eq314}
\mathcal{N}_{n_{\parallel}} = \sqrt{2n_{\parallel} +
\gamma + \beta + 1}\, \sqrt{\frac{\Gamma(n_{\parallel} +
1)\Gamma(n_{\parallel} + \gamma + \beta + 1)} {\Gamma(n_{\parallel}
+ \gamma + 1)\Gamma(n_{\parallel} + \beta + 1)}}.
\end{equation}

Including both transverse and longitudinal contributions, the total
meson mass squared becomes
\begin{equation}\label{eq315}
m^2(n_{\perp}, n_{\parallel}, L, S) = 2\kappa^2
(2n_{\perp} + 2L + S) + \sigma (2n_{\parallel} + 1)(m_{q} + m_{\bar
q}) + \sigma^2 n_{\parallel} (n_{\parallel} + 1) + (m_q + m_{\bar
q})^2.
\end{equation}

The corresponding quark-antiquark wave function reads
\begin{equation}\label{eq316}
\psi^{\rm LMZV}_{n_\perp,n_\parallel} = \psi^{\rm
LFH}_{n_\perp}(\zeta,\varphi,\alpha)\, \chi^{\rm
LMZV}_{n_\parallel}(\alpha).
\end{equation}

It is worth emphasizing that the LMZV construction provides a fully
analytic realization of longitudinal confinement while preserving
the endpoint behavior and chiral properties encoded in the 't Hooft
framework. In contrast to the numerical diagonalization required in
the 't Hooft model, the LMZV approach yields closed-form expressions
for both eigenvalues and eigenfunctions, significantly simplifying
phenomenological applications. This analytic control facilitates
systematic investigations of hadronic observables while maintaining
consistency with large-$N_c$ QCD dynamics.

\section{Distribution Amplitudes of the $K_0^*$ meson}
In this section, we derive the twist-$2$ and twist-$3$ DAs of the
$K_0^*(1430)$ meson within the framework of HLFQCD, starting from
the corresponding LFWF. The light-cone DAs of $K_0^*(1430)$,
composed of a strange quark $s$ and a light antiquark $\bar q$
($\bar u$ or $\bar d$), are defined through the meson-to-vacuum
matrix elements
\begin{align}
\langle 0 | \bar q(y)\gamma^{\mu} s(x)|K_0^*(p)\rangle&=
\,f_{_{K_0^*}} p^{\mu} \int_0^1 d\alpha\,e^{-i(\alpha p\cdot x+\bar \alpha p\cdot y)} \Phi(\alpha, \mu)\,, \label{eq401}\\
\langle 0 | \bar q(y)\, s(x)|K_0^*(p)\rangle&=
m_{_{K_0^*}} \bar f_{_{K_0^*}} \int_0^1 d\alpha\, e^{-i(\alpha p\cdot x+\bar \alpha p\cdot y)} \Phi_s(\alpha, \mu)\,, \label{eq402}\\
\langle 0 | \bar q(y) \sigma^{\mu\nu} s(x)|K_0^*(p)\rangle&=
-\frac{m_{_{K_0^*}}}{6}(p^\mu z^\nu - p^\nu z^\mu)\bar f_{_{K_0^*}}
\int_0^1 d\alpha\, e^{-i(\alpha p\cdot x+\bar \alpha p\cdot y)}
\Phi_{\sigma}(\alpha, \mu)\,, \label{eq403}
\end{align}
where $z=y-x$. Here $m_{K_0^*}$ and $p^\mu$ denote the mass and
momentum of the meson, while $f_{K_0^*}$ and $\bar f_{K_0^*}$ are
the vector-current and scalar-current decay constants, respectively.
The function $\Phi(\alpha,\mu)$ represents the leading-twist
(twist-$2$) DA, whereas $\Phi_s(\alpha,\mu)$ and
$\Phi_\sigma(\alpha,\mu)$ correspond to the twist-$3$ DAs. The
parameter $\mu$ denotes the renormalization scale.

In the light-front coordinate system, the space-time four-vector and
the meson four-momentum are defined as
\begin{equation}\label{eq404}
x^{\mu}=(x^+,x^-,\mathbf{x}_\perp), \qquad
p^{\mu}=\Big(p^+,\frac{m_{K_0^*}^2}{p^+},\mathbf{0}_\perp\Big).
\end{equation}
Choosing the light-front time $x^+=0$ and setting $y^\mu=0$ in Eqs.
(\ref{eq401})--(\ref{eq403}), one obtains
\begin{align}
\langle 0 | \bar q(0)\gamma^{\mu} s(x)|K_0^*(p)\rangle&=
\,f_{_{K_0^*}} p^{\mu} \int_0^1 d\alpha\,e^{-i\alpha p^+ x^- }\Phi(\alpha, \mu)\,,\label{eq404}\\
\langle 0 | \bar q(0)~ s(x)|K_0^*(p)\rangle&=
m_{_{K_0^*}} \bar f_{_{K_0^*}} \int_0^1 d\alpha\, e^{-i\alpha p^+ x^- } \Phi_s(\alpha, \mu)\,,\label{eq405}\\
\langle 0 | \bar q(0) \sigma^{\mu\nu} s(x)|K_0^*(p)\rangle&=
\frac{m_{_{K_0^*}}}{6} (p^\mu x^\nu - p^\nu x^\mu)\bar f_{_{K_0^*}}
\int_0^1 d\alpha\,e^{-i\alpha p^+ x^- } \Phi_{\sigma}(\alpha,
\mu)\,.\label{eq406}
\end{align}
To isolate the leading-twist and scalar twist-$3$ DAs, we choose the
``$+$'' component ($\mu=+$) in Eqs. (\ref{eq404}) and (\ref{eq405}).
Performing the Fourier transform with respect to the longitudinal
distance $x^-$ yields
\begin{align}
\Phi(\alpha, \mu)&= \frac{1}{f_{_{K_0^*}}}\int dx^{-}\,e^{i \alpha
p^{+} x^{-}} \langle 0 | \bar q(0)\gamma^{+} s(x)|K_0^*(p)\rangle
,\label{eq407}\\
\Phi_{s}(\alpha, \mu)&=\frac{p^{+} }{ m_{_{K_0^*}} \bar
f_{_{K_0^*}}} \int dx^{-}\,e^{i \alpha p^{+} x^{-}} \langle0 | \bar
q(0) ~s(x)|K_0^*(p)\rangle.\label{eq408}
\end{align}
For the tensor DA, selecting $\sigma^{+-}$ in Eq. (\ref{eq406}) and
performing integration by parts, together with the boundary
condition $\Phi_{\sigma}(\alpha)\big|_0^1=0$, one obtains after
Fourier transformation
\begin{equation}\label{eq409}
\frac{\partial \Phi_{\sigma}(\alpha, \mu)}{\partial \alpha}=\frac{6
p^{+}} {i m_{_{K_0^*}} \bar f_{_{K_0^*}}}  \int dx^{-} e^{i \alpha
p^{+} x^{-}} \langle 0 | \bar q(0) \sigma^{+ -}
s(x)|K_0^*(p)\rangle.
\end{equation}

To determine the holographic DAs of the $K_0^*$ meson, the hadronic
matrix elements in Eqs. (\ref{eq407})--(\ref{eq409}) must be
evaluated using the corresponding LFWF. The Fock-state expansion of
the meson eigenstate $|K_0^*(p)\rangle$ in terms of free
two-particle states reads \cite{ForSan}
\begin{equation}\label{eq410}
|K_0^*(p)\rangle=\sum_{h,\bar{h}} \int \frac{d k^{+} d^2{\bf
{k}_{\bot}}}{16\pi^3 \sqrt{k^{+}(p^{+}-k^{+})}} {\Psi}_{h,\bar{h}} (
\alpha, {\bf {k}_{\bot}} ) |\alpha p^+,\textbf{k}_{\bot},h;\,{\bar
\alpha} p^+,-\textbf{k}_{\bot},\bar{h}\rangle,
\end{equation}
where $k^+=\alpha p^+$ is the longitudinal momentum of the strange
quark, and $\Psi_{h\bar h}(\alpha,\mathbf{k}_{\perp})$ denotes the
LFWF.

The matrix element $\langle 0 | \bar q(0)\,\Gamma\, s(x^-)
|K_0^*(p)\rangle$, with $\Gamma = I,\gamma^+$, or $\sigma^{+-}$, is
then obtained as
\begin{align}\label{eq411}
\langle 0 | \bar{q}(0)\, \Gamma\, s(x^-) |K_0^*(p) \rangle&= \sqrt{
N_c} \sum_{h,\bar{h}} \int \frac{d k^{+}d^2{\bf {k}_{\bot}}
\,\Theta(|\mathbf{k}_{\bot}| <\mu)}
{16\,\pi^3\sqrt{k^+(p^{+}-k^{+})}}
\Psi_{h,\bar{h}}( \alpha, {\bf {k}_{\bot}} )\nonumber \\
& \times \bar{q}_{\bar{h}}(\bar{\alpha} p^+,-{\bf {k}_{\bot}})\,
\Gamma\, s_h(\alpha p^+,{\bf {k}_{\bot}}) e^{-ik^{+}x^{-}}\,,
\end{align}
here, $s_h$ and $q_{\bar h}$ denote light-front helicity spinors.
Equation \eqref{eq411} follows from the light-front mode expansion
of the Dirac fields together with the equal light-front time
anticommutation relations. The parameter $\mu$ serves as the
renormalization scale and provides an ultraviolet cutoff on the
transverse momenta \cite{KogSus,Diehl}. This scale enters through
the constraint $\Theta(|\mathbf{k}_\perp| < \mu)$, ensuring
consistency between the light-front formalism and the scale
dependence of QCD distribution amplitudes. Upon performing the $k^+$
integration and applying the Fourier transform with respect to
$x^-$, one obtains
\begin{align}\label{eq412}
\int d x^- e^{i\alpha p^+x^-} \langle 0 | \bar{q}(0)\,  \Gamma\,
s(x^-)|K_0^*(p) \rangle &= \frac{\sqrt{ N_c}}{p^+}\sum_{h,\bar{h}}
\int^{{k}_{\bot} < \mu}
\frac{d^2\mathbf{k}_{\bot}}{(2\pi)^3}\Psi_{h,\bar{h}}(\alpha,\mathbf{k}_{\bot})\\
\nonumber &\times \left \{ \frac{\bar{q}_{\bar{h}}(\bar \alpha
p^{+},-\mathbf{k}_{\bot})}{\sqrt{\bar \alpha}} \,\Gamma\,
\frac{s_h(\alpha p^+,\mathbf{k}_{\bot})}{\sqrt{\alpha}} \right\}\,.
\end{align}
Using the light-front helicity spinors given in Eq. (\ref{eq204}),
the spinor structures appearing on the right-hand side of Eq.
(\ref{eq412}) are evaluated as
\begin{align}
\frac{\bar{q}_{\bar{h}}}{\sqrt{\bar{\alpha}}}\gamma^+\frac{s_h}{\sqrt{\alpha}} &=2 p^+ \delta_{h,\pm} \delta_{\bar{h},\mp}\,,\label{eq413}\\
\frac{\bar{q}_{\bar{h}}}{\sqrt{\bar{\alpha}}}\ I \frac{s_h}{\sqrt{\alpha}} &=\frac{1}{\alpha \bar \alpha}[k_{\bot} e^{\pm i\theta} \delta_{h,\pm}\delta_{\bar{h},\pm}+\mathcal{\bar{M}}_{-}\,\delta_{h,\pm}\delta_{\bar{h},\mp}]\,,\label{eq414}\\
\frac{\bar{q}_{\bar{h}}}{\sqrt{\bar{\alpha}}}\sigma^{+-}\frac{s_h}{\sqrt{\alpha}}
&=\frac{2i}{\alpha \bar \alpha}[(\mp \alpha \pm \bar \alpha)k_{\bot}
e^{\pm i\theta}
\delta_{h,\pm}\delta_{\bar{h},\pm}+\mathcal{\bar{M}}_{+}\,\delta_{h,\pm}\delta_{\bar{h},\mp}]\,.\label{eq415}
\end{align}

Using Eqs. (\ref{eq413})--(\ref{eq415}) and inserting the spin wave
function $S_{h\bar h}$ from Eq. (\ref{eq206}), the DAs $\Phi$,
$\Phi_s$, and $\partial_\alpha \Phi_\sigma$ in Eqs.
(\ref{eq407})--(\ref{eq409}) can be expressed in terms of the
quark-antiquark wave function as
\begin{align}
\Phi(\alpha, \mu) &= \frac{2 N \sqrt{N_{c}}}{{\alpha
\bar{\alpha}}\,\pi\,f_{_{K^{*}_0}}}\, \int^{{k}_{\bot} < \mu}
\frac{d^2 \bf {k}_{\bot}}{4\pi^2}  \left\{
\mathcal{M}^2-{k}_{\bot}^{2}+
m_{_{K^{*}_0}}\mathcal{\bar{M}}_{-}\right\} \psi(\alpha, \bf
{k}_{\bot})\,,\label{eq416}
\\
\Phi_s(\alpha, \mu) &= \frac{N \sqrt{N_{c}}}{{\alpha^2
{\bar{\alpha}}^2}\,\pi\,{\bar{f}}_{_{K^{*}_0}}\,m_{_{K^{*}_0}}}\,
\int^{k_{\bot} < \mu} \frac{d^2 \bf {k}_{\bot}}{4\pi^2}
\left\{\mathcal{M}_{+}\,{k}^{2}_{\bot}+(\mathcal{M}^2
-{k}_{\bot}^{2})\,\mathcal{\bar{M}}_{-}+
m_{_{K^{*}_0}}\mathcal{\bar{M}}^{2}_{-}\right\} \psi(\alpha, \bf
{k}_{\bot})\,,\label{eq417}
\\
\frac{\partial \Phi_{\sigma}(\alpha, \mu)}{\partial \alpha}&=
\frac{12 N\,\sqrt{N_{c}}}{{\alpha^2
{\bar{\alpha}}^2}\,\pi\,{\bar{f}}_{_{K^{*}_0}}\,m_{_{K^{*}_0}}}\,
\int^{{k}_{\bot} < \mu} \frac{d^2 \bf {k}_{\bot}}{4\pi^2}
\left\{\mathcal{M}_{+}\,{k}^{2}_{\bot}(1-2\alpha) + (\mathcal{M}^2
-{k}_{\bot}^{2})\,\mathcal{\bar{M}}_{+}+
m_{_{K^{*}_0}}\mathcal{\bar{M}}_{+}\mathcal{\bar{M}}_{-} \right\}
\psi(\alpha, \bf {k}_{\bot}).\label{eq418}
\end{align}

With the normalization conditions
\begin{equation}\label{eq419}
\int_0^1 d\alpha\, \Phi(\alpha,\mu)=1, \qquad \int_0^1 d\alpha\,
\Phi_s(\alpha,\mu)=1,
\end{equation}
the decay constants are defined through the local limits
\begin{equation}\label{eq420}
\langle 0|\bar q \gamma^\mu s|K_0^*(p)\rangle = p^\mu f_{K_0^*},
\qquad \langle 0|\bar q s|K_0^*(p)\rangle = m_{K_0^*}\bar f_{K_0^*}.
\end{equation}

Following the same light-front procedure, the vector decay constant
is obtained as
\begin{equation}\label{eq421}
f_{_{K^{*}_0}} = \frac{2 N \sqrt{N_{c}}}{\pi}\int_{0}^{1} {d \alpha
} \int \frac{d^2 \bf {k}_{\bot}}{4\pi^2}  \left\{
\mathcal{M}^2-{k}_{\bot}^{2}+
m_{_{K^{*}_0}}\mathcal{\bar{M}}_{-}\right\} \frac{\psi(\alpha, \bf
{k}_{\bot})}{\alpha \bar{\alpha}}.
\end{equation}

The vector decay constant $f_{_{K_0^*}}$ and the scale-dependent
scalar decay constant $\bar f_{_{K_0^*}}$ are related through the
equation of motion \cite{CDLu,MZYang}
\begin{equation}\label{eq422}
{\bar {f}}_{_{K^{*}_0}} =
\frac{m_{_{K^{*}_0}}}{m_s-m_{\bar{q}}}f_{_{K^{*}_0}}.
\end{equation}

\section{NUMERICAL ANALYSIS}
In this section, we present a comprehensive numerical analysis based
on the three forms of the LFWF constructed within the IMA, modified
't Hooft, and LMZV frameworks. These approaches represent different
realizations of longitudinal confinement and chiral symmetry
breaking. In particular, the IMA model employs a phenomenological
invariant-mass ansatz for the longitudinal dynamics, whereas the
modified 't Hooft and LMZV frameworks determine the longitudinal
wave function by solving dynamical longitudinal equations. The
resulting wave functions provide the basis for calculating the decay
constants, distribution amplitudes, transition form factors, and
other physical observables associated with the scalar meson
$K_0^*(1430)$.

To ensure a meaningful comparison among the three longitudinal
dynamical models, the same set of model parameters listed in Table
\ref{T301} is employed throughout the IMA, LMZV, and 't Hooft
models. This choice is made deliberately to perform a controlled
comparison of the different implementations of longitudinal
confinement. By employing identical input parameters, the observed
differences in the predicted observables can be attributed primarily
to the different implementations of longitudinal confinement rather
than to differences in the model parameters. In contrast, adopting
different parameter sets for different models would introduce
additional model-dependent degrees of freedom, making it difficult
to disentangle the effects of longitudinal dynamics from those
arising solely from parameter choices.

We compute the decay constants, holographic DAs and their
$\xi$-moments, and valence PDF. Furthermore, using the twist-$2$ and
twist-$3$ DAs, we evaluate the transition form factors, lepton
polarization asymmetries, and differential branching ratios for the
semileptonic decay $B \to K_0^* \ell^+ \ell^-$, with $\ell = \mu,
\tau$, as functions of the momentum transfer $q^2$. This broad set
of observables allows us to systematically examine the sensitivity
of both static and dynamical quantities to the underlying
longitudinal structure of the LFWF.

The LFWF of mesons is characterized by the longitudinal and
transverse quantum numbers $n_\parallel$ and $n_\perp$. In Table
\ref{T501}, these quantum numbers are assigned for the kaon family
such that, together with the corresponding orbital and spin quantum
numbers, the resulting mass spectra satisfy the mass relations
derived within the 't Hooft and LMZV frameworks. The predicted
masses obtained in both longitudinal approaches are compared with
the available experimental data reported by the Particle Data Group
(PDG) \cite{PDG}, which are also listed in the table for direct
comparison.
\begin{table}[ht]
\caption{Quantum numbers and mass spectrum of the $K$-meson family,
including theoretical predictions from the 't Hooft and LMZV models,
compared to the experimental values from PDG \cite{PDG}.}
\label{T501}
\begin{ruledtabular}
\begin{tabular}{ccccccccc|ccccccccc}
textrm{Meson} & $J^{P}$ & $S$ & $L$ & $n_{\perp}$ & $n_{\parallel}$
&  $M_{\text{'t Hooft}}$ & $M_{\text{LMZV}}$  & $M_{\text{PDG}}$ &
\textrm{Meson} & $J^{P}$ & $S$ & $L$ & $n_{\perp}$ & $n_{\parallel}$ &  $M_{\text{'t Hooft}}$ & $M_{\text{LMZV}}$  & $M_{\text{PDG}}$ \\
\hline
$K(493)$        & $0^-$ & 0 & 0 & 0 & 0 & $636 \pm 29$    & $646 \pm 34$   &  $459^{+47}_{-45}$            & $K^*(892)$      & $1^-$ & 1 & 0 & 0 & 0 & $977 \pm 25$    & $983 \pm 33$   & $892$    \\[2mm]
$K(1460)$       & $0^-$ & 0 & 0 & 1 & 1 & $1337 \pm 45$   & $1387 \pm 49$  &  $1460$                       & $K^*(1410)$     & $1^-$ & 1 & 0 & 1 & 1 & $1468 \pm 49$   & $1495 \pm 51$  & $1410$   \\[2mm]
                &       &   &   &   &   &                 &                &                               & $K^*(1680)$     & $1^-$ & 1 & 0 & 2 & 2 & $1800 \pm 61$   & $1890 \pm 63$  & $1680$   \\[2mm]
\hline
$K_1(1270)$     & $1^+$ & 0 & 1 & 0 & 1 & $1337 \pm 45$   & $1387 \pm 49$  & $1270$                        & $K_0^*(1430)$   & $0^+$ & 1 & 1 & 0 & 0 & $1432 \pm 50$   & $1437 \pm 53$  & $1430$   \\[2mm]
$K_1(1400)$     & $1^+$ & 0 & 1 & 0 & 2 & $1381 \pm 45$   & $1463 \pm 54$  & $1400$                        & $K_2^*(1430)$   & $2^+$ & 1 & 1 & 0 & 1 & $1477 \pm 47$   & $1539 \pm 51$  & $1430$   \\[2mm]
$K_1(1650)$     & $1^+$ & 0 & 1 & 1 & 2 & $1687 \pm 58$   & $1773 \pm 61$  & $1650$                        & $K_0^*(1950)$   & $0^+$ & 1 & 1 & 1 & 2 & $1826 \pm 66$   & $1879 \pm 66$  & $1950$   \\[2mm]
                &                &   &   &   &   &        &                &                               & $K_2^*(1980)$   & $2^+$ & 1 & 1 & 2 & 3 & $2081 \pm 66$   & $2207 \pm 71$  & $1980$   \\[2mm]
\hline
$K_2(1770)$     & $2^-$ & 0 & 2 & 0 & 2 & $1687 \pm 58$   & $1773 \pm 61$  & $1770$                        & $K_3^*(1780)$   & $3^-$ & 1 & 2 & 0 & 2 & $1800 \pm 61$   & $1890 \pm 63$  & $1780$   \\[2mm]
$K_2(1820)$     & $2^-$ & 0 & 2 & 1 & 3 & $1937 \pm 63$   & $2051 \pm 68$  & $1820$                        &                 &       &   &   &   &   &                 &                &          \\[2mm]
\end{tabular}
\end{ruledtabular}
\end{table}
In particular, the scalar meson $K_0^*(1430)$ is identified with the
quantum numbers $(S, L, n_\perp, n_\parallel) = (1, 1, 0, 0)$,
leading to predicted masses of $1432 \pm 50~\mathrm{MeV}$ and $1437
\pm 53~\mathrm{MeV}$ in the 't Hooft and LMZV models, respectively.
These values are in close agreement with the experimental
measurement and lie well within the quoted PDG uncertainties,
indicating that both longitudinal frameworks provide a
quantitatively consistent description of the kaon mass spectrum.

\begin{figure}[h]
\begin{center}
\includegraphics[width=8.75cm,height=8cm]{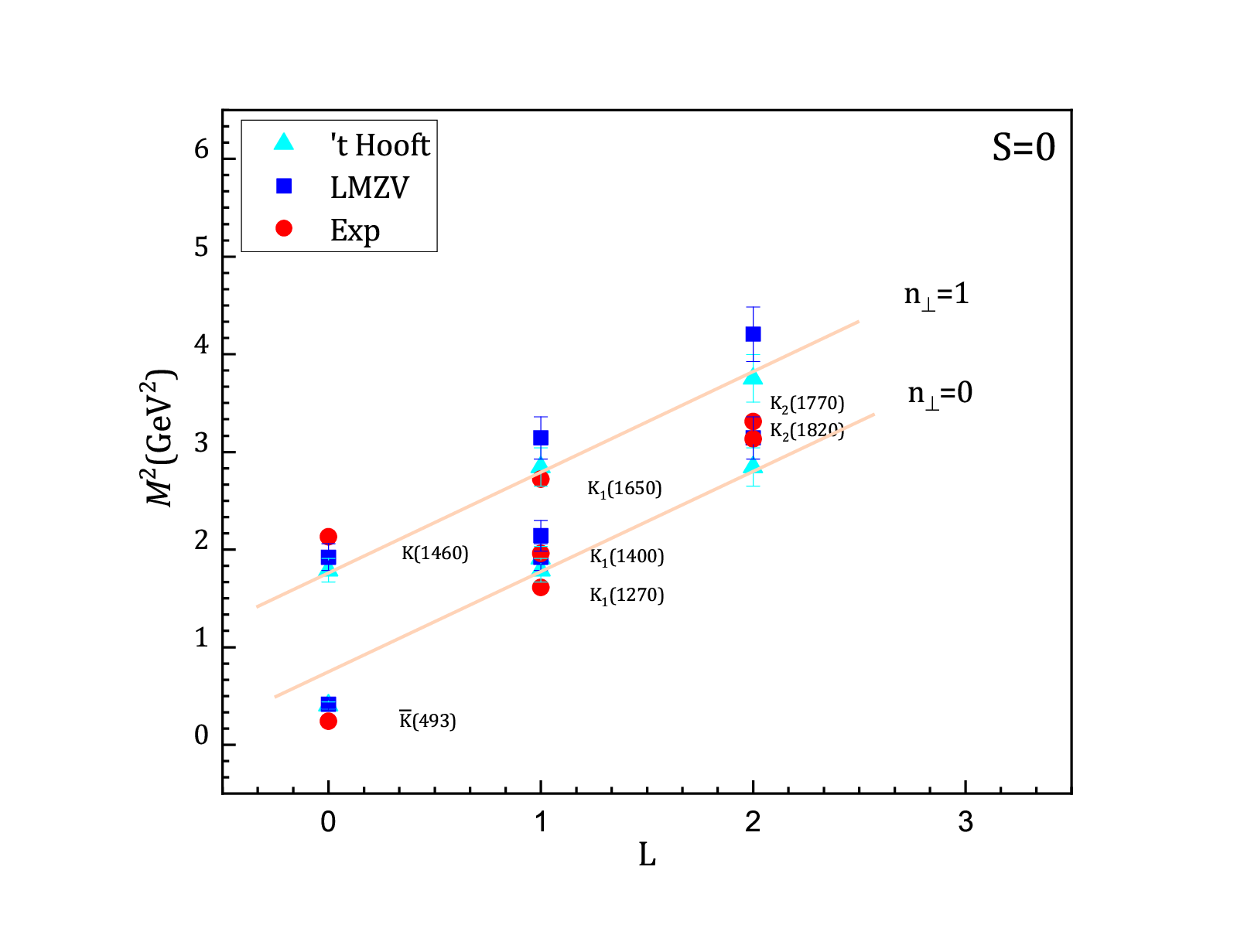}
\includegraphics[width=8.75cm,height=8cm]{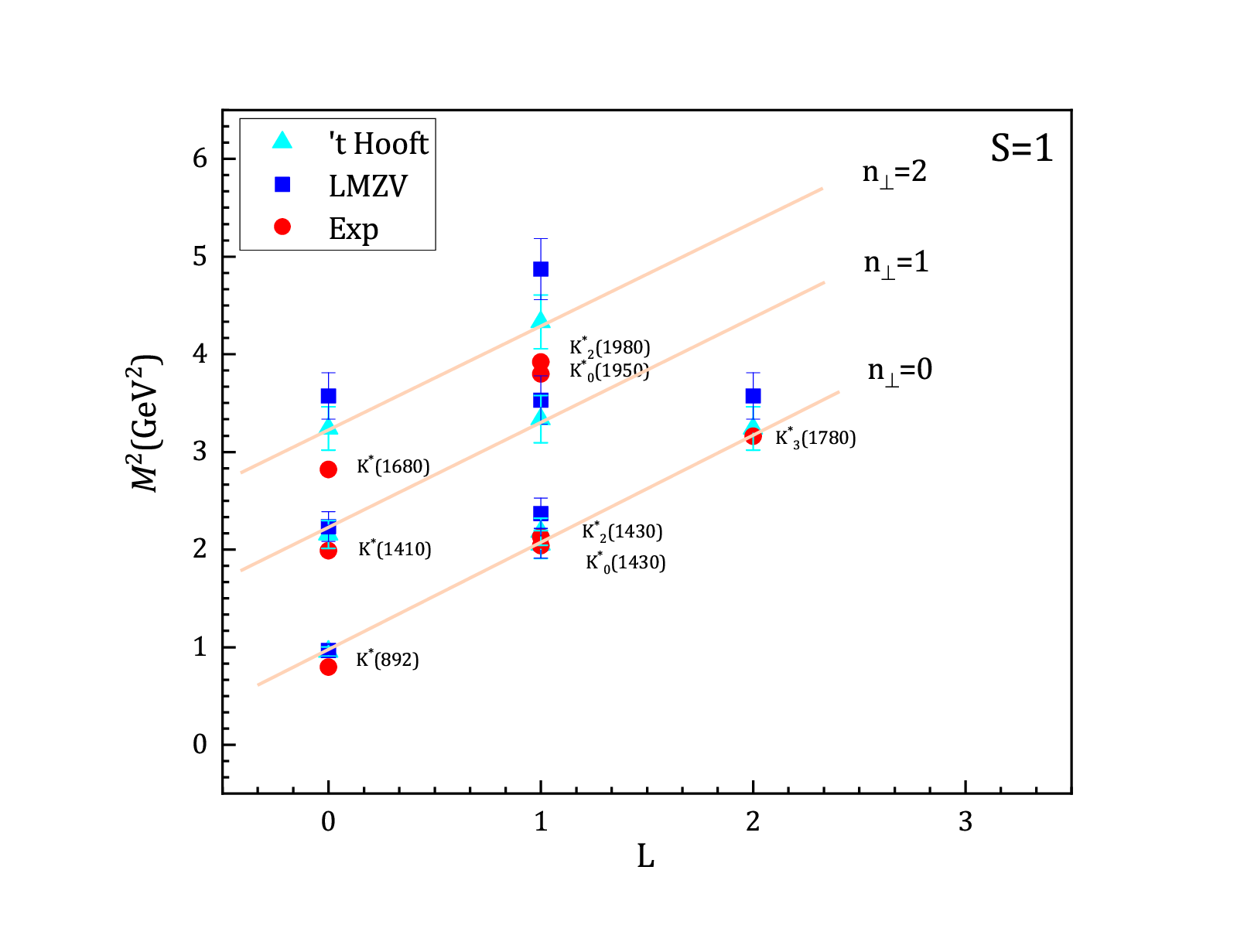}
\caption{The mass squared on the orbital angular momentum for the
kaon family.} \label{F501}
\end{center}
\end{figure}

The mass trajectories of the kaon family are further illustrated in
Fig. \ref{F501}, where the squared masses are plotted as functions
of the orbital angular momentum $L$. Both models exhibit
approximately linear Regge-like behavior, reflecting the confinement
dynamics encoded in the light-front Hamiltonian. For definiteness,
the Regge trajectories shown in the figure are constructed using the
't Hooft masses. The near-linearity of these trajectories supports
the adopted assignment of quantum numbers and provides a consistent
foundation for the subsequent computation of observables based on
the LFWF.

The normalization constant $N$ appearing in Eq. (\ref{eq201}) is
fixed by the probabilistic normalization condition of the LFWF,
which was already introduced in Eq. (\ref{eq202}). The resulting
values of the normalization constant obtained within the IMA, 't
Hooft, and LMZV approaches are summarized in Table \ref{T502}.
\begin{table}[ht]
\caption{Normalization constants of the wave functions for the IMA,
't Hooft , and LMZV methods.} \label{T502}
\begin{ruledtabular}
\begin{tabular}{cccc}
Approach                & IMA      & 't Hooft     & LMZV    \\
\hline $N$                & $2.33^{+0.83}_{-0.68}$  &   $
0.91^{+0.07}_{-0.07} $   & $0.87^{+0.07}_{-0.07}$
\end{tabular}
\end{ruledtabular}
\end{table}

\begin{figure}[h]
\begin{center}
\includegraphics[width=9.75cm,height=8cm]{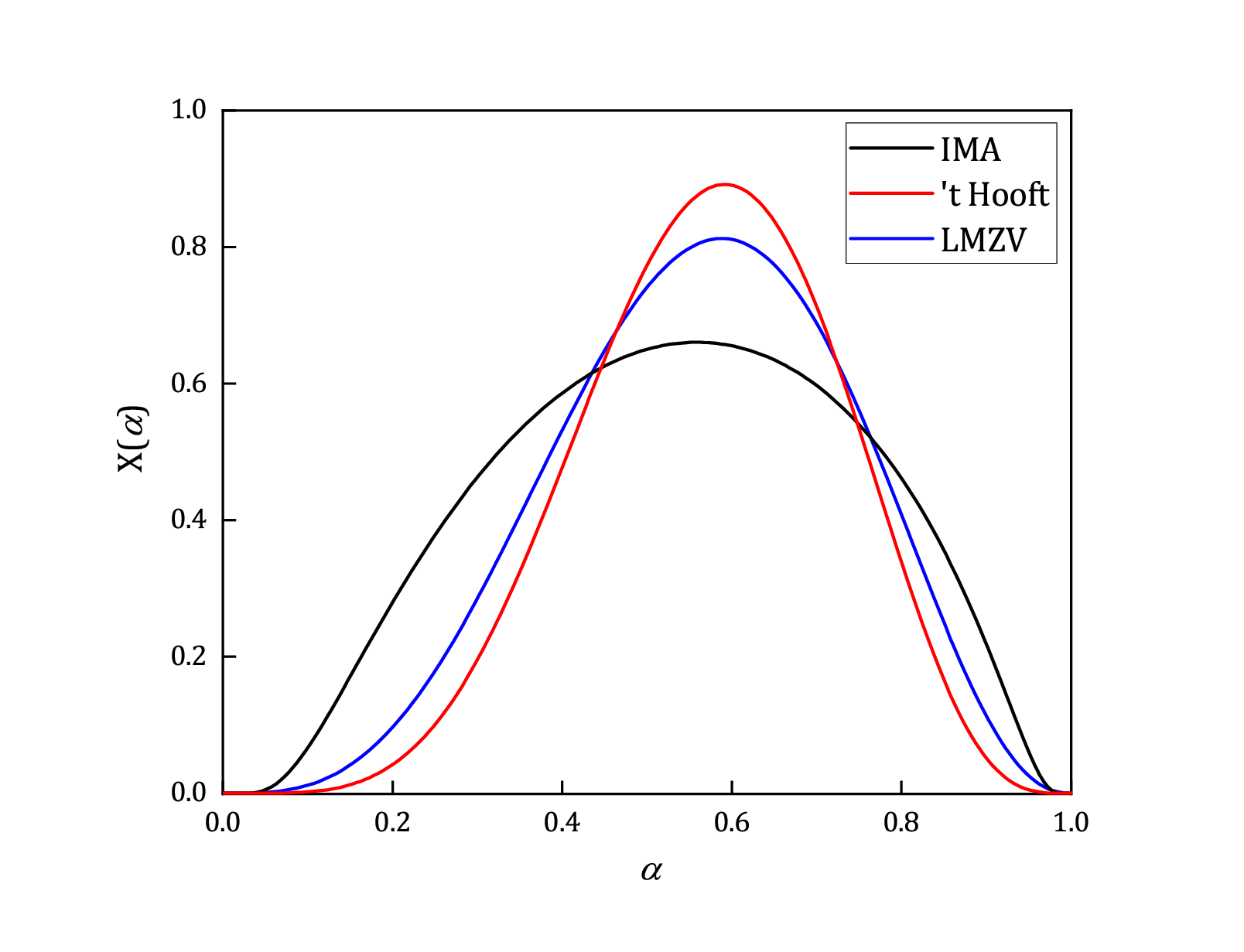}
\caption{The longitudinal component $X(\alpha)$ of the LFWF for the
scalar meson obtained in the IMA (black), 't Hooft (red), and LMZV
(blue) approaches.} \label{F5010}
\end{center}
\end{figure}

The longitudinal LFWF $X(\alpha)$ for the $K_0^*(1430)$ meson,
computed within the IMA, 't Hooft, and LMZV frameworks, is shown in
Fig. \ref{F5010}.
This quantity encodes the distribution of the longitudinal momentum
fraction $\alpha$ between the quark and antiquark constituents and
therefore provides direct insight into the internal light-front
dynamics of the system. All three approaches exhibit the expected
physical endpoint behavior, $X(\alpha)\to 0$ as $\alpha\to 0, 1$,
ensuring the suppression of configurations in which a single
constituent carries either vanishing or nearly all of the
longitudinal momentum.

Despite this common feature, the three models lead to distinct
shapes for $X(\alpha)$, reflecting differences in their underlying
dynamical mechanisms. The IMA prediction yields the broadest
distribution, characteristic of its phenomenological structure and
the relatively mild constraints it imposes on the momentum
partitioning. In contrast, the 't Hooft model produces a noticeably
sharper peak centered near $\alpha \approx 0.5$, consistent with the
confining kernel of the effective $(1+1)$ dimensional QCD
interaction that drives the distribution toward more symmetric
quark-antiquark momentum sharing. The LMZV result lies between these
two limits, exhibiting a smoother profile with a peak lower than
that of the 't Hooft model while remaining less diffuse than the IMA
curve.

Taken together, the three predictions illustrate how different
dynamical assumptions- ranging from phenomenological modeling to
more tightly constrained effective QCD kernels- manifest in the
longitudinal structure of the LFWF. These distinctions directly
influence observables derived from the wave function, such as
$\xi$-moments and DAs, thereby providing a useful benchmark for
assessing the impact of model-dependent dynamics on the internal
structure of the $K_0^*$ meson.

The decay constants $f_{_{K_0^*}}$ and $\bar{f}_{_{K_0^*}}$ are
calculated using Eqs. (\ref{eq421}) and (\ref{eq422}) at the scale
$\mu=1$ GeV, and the results are presented in Table \ref{T503}.
\begin{table}[h]
\caption{Predicted values for the decay constants $\bar{f}_{K_0^*}$
and $f_{K_0^*}$ from different models (in MeV).} \label{T503}
\begin{ruledtabular}
\begin{tabular}{cccc}
Approach                & $ f_{_{K_0^*}}$(MeV)& Approach            & $\bar{f}_{_{K_0^*}}$ (MeV)    \\
\hline
IMA                     & $ 28^{+15}_{-15} $& IMA                  & $ 285^{+43}_{-37}  $  \\
't Hooft                & $ 37^{+11}_{-11} $& 't Hooft             & $ 428^{+63}_{-54} $  \\
LMZV                    & $ 39^{+11}_{-11} $& LMZV                 & $ 444^{+65}_{-56} $  \\
LFQM \cite{ChGeLi}      & $ 21 \pm 5     $& QCDSR \cite{HanWuFu} & $ 358             $  \\
QCDSR \cite{KMaltman}   & $ 42 \pm 2     $& PQCD \cite{RunHui}   & $ 445 \pm 50      $  \\
QCDSR \cite{VChernyak}  & $ 70 \pm 10    $& QCDSR \cite{CDLu}    & $ 440 \pm 15      $  \\
CLF \cite{ChZhZha}      & $ 43 \pm 10    $& QCDSR \cite{DuLiYa}  & $ 427 \pm 85      $  \\
NJL \cite{CMShakin}     & $ 31           $&                      &                      \\
\end{tabular}
\end{ruledtabular}
\end{table}
The predictions obtained within the IMA, 't Hooft, and LMZV
frameworks are compared with several other theoretical approaches
and, overall, show reasonable agreement with the values reported in
the literature. For $f_{_{K_0^*}}$, the results of these three
models lie within the range of values reported by QCDSR
\cite{KMaltman,VChernyak}, the covariant light-front approach (CLF)
\cite{ChZhZha}, LFQM \cite{ChGeLi}, and the Nambu-Jona-Lasinio (NJL)
model \cite{CMShakin}. In particular, the predictions of the 't
Hooft and LMZV models are numerically closer to QCDSR
\cite{KMaltman} and the CLF results \cite{ChZhZha}, which may be
attributed to the explicit implementation of longitudinal
confinement dynamics in these frameworks.

As seen in Table \ref{T503}, the decay constants predicted in the 't
Hooft and LMZV approaches are very close to each other. This
proximity can be traced back to the common dynamical origin of the
two frameworks. In the 't Hooft formalism, the longitudinal
potential arises from an integral equation associated with chiral
symmetry breaking and possesses a nonlocal structure that governs
the longitudinal dynamics of quarks in the scalar channel. In the
LMZV approach, an analytic and solvable potential is introduced that
reproduces many of the essential dynamical features of the 't Hooft
model, including the endpoint behavior, chiral dependence, and its
effective nonlocal character. This similarity in the underlying
dynamics naturally explains the strong numerical agreement observed
between the two models. By contrast, the prediction of the IMA model
is closer to the result obtained in the NJL model \cite{CMShakin},
which is consistent with the more phenomenological nature of this
model in describing the longitudinal structure.

For the scalar decay constant $\bar{f}_{_{K_0^*}}$, our results are
compared with PQCD predictions \cite{RunHui} and several QCDSR
analyses \cite{CDLu,DuLiYa,HanWuFu}. In this case as well, the
values obtained from the 't Hooft and LMZV models remain close to
each other and are consistent, within uncertainties, with PQCD
results and some QCDSR determinations \cite{CDLu,DuLiYa}. The value
obtained in the IMA model is somewhat smaller and shows a larger
deviation from these results, indicating that $\bar{f}_{_{K_0^*}}$
is sensitive to the specific modeling of the longitudinal dynamics.
Overall, the pattern observed in Table \ref{T503} suggests that the
inclusion of dynamical longitudinal confinement plays an important
role in stabilizing the value of the scalar decay constant.

\begin{figure}[h]
\begin{center}
\includegraphics[width=5.5cm,height=5cm]{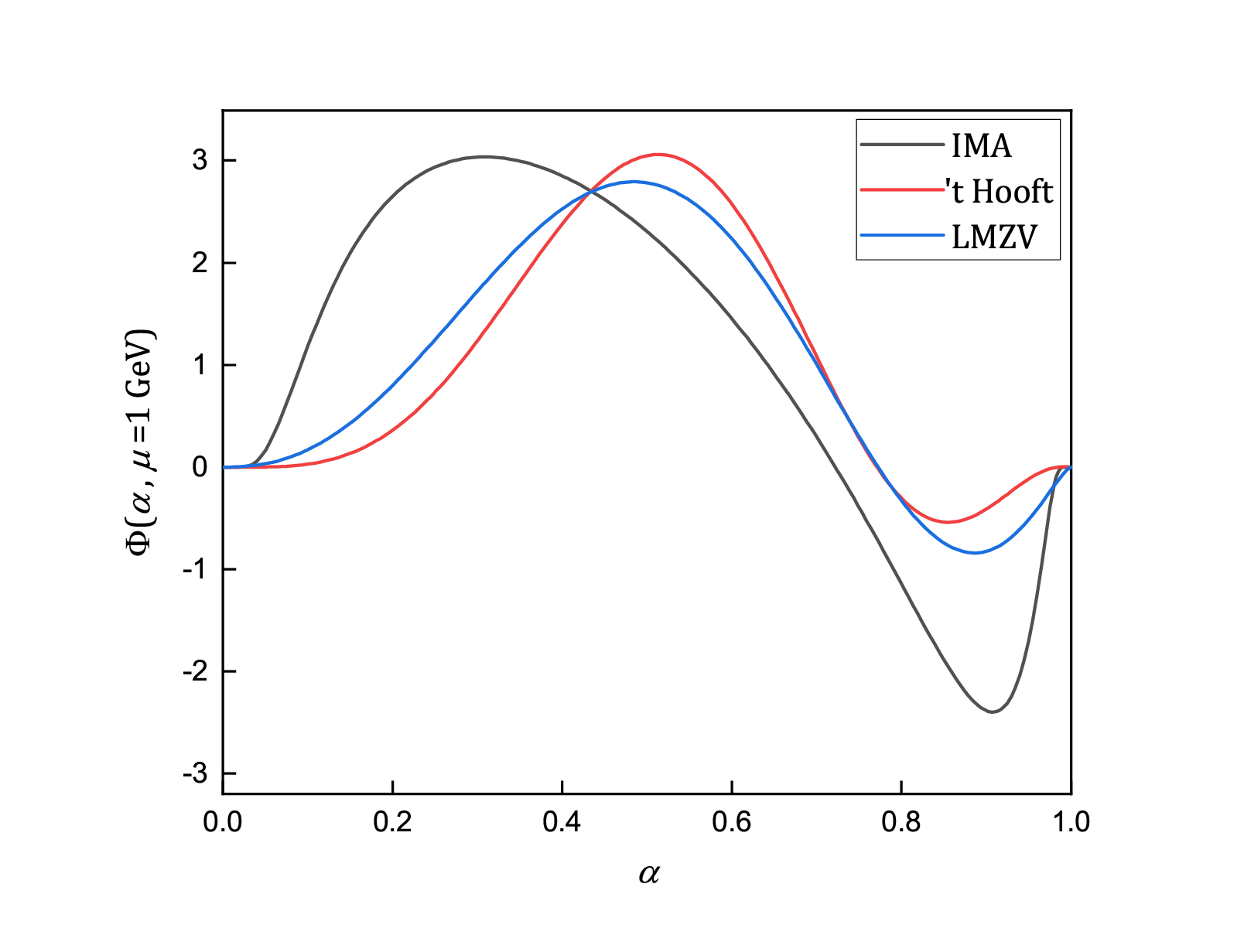}
\includegraphics[width=5.5cm,height=5cm]{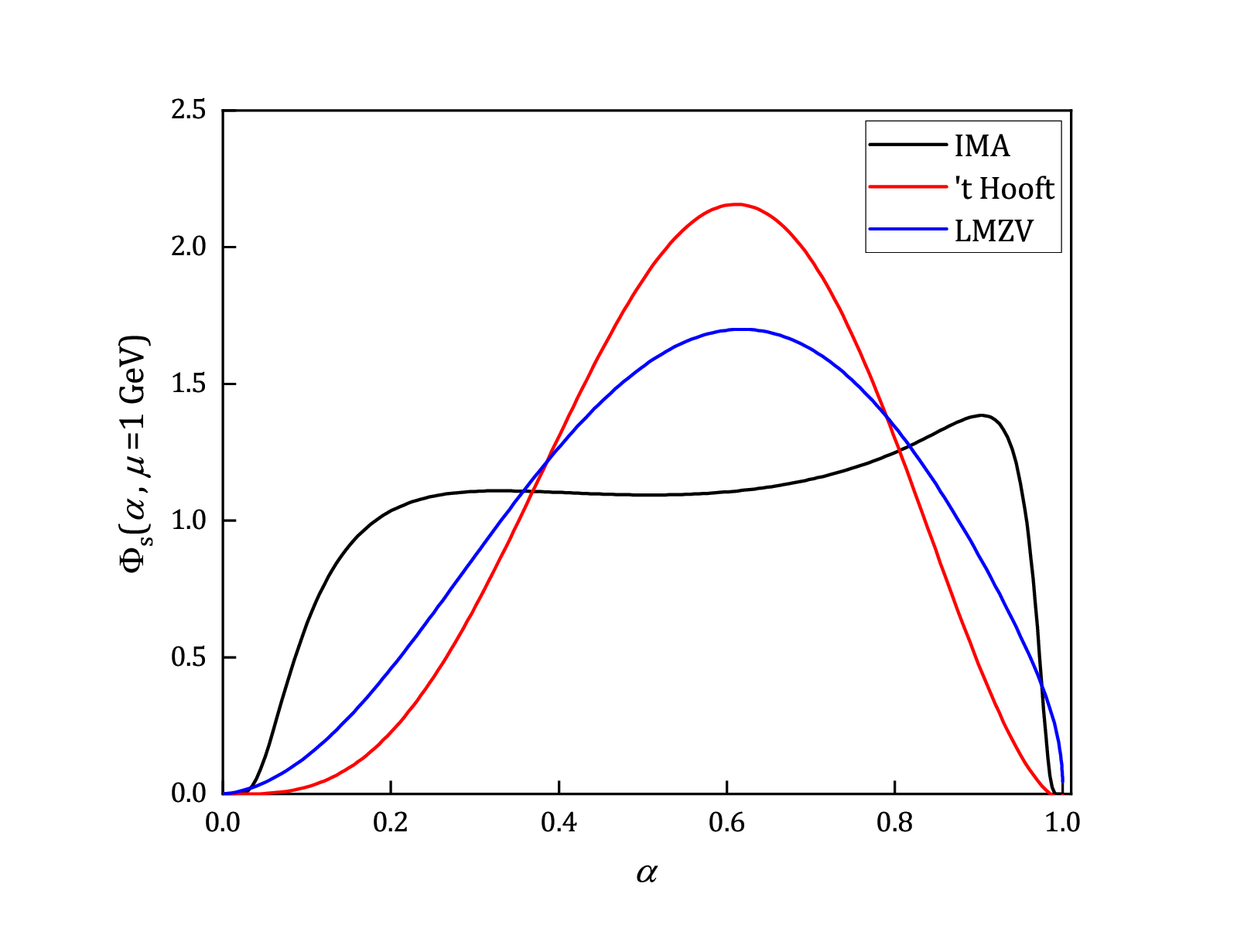}
\includegraphics[width=5.5cm,height=5cm]{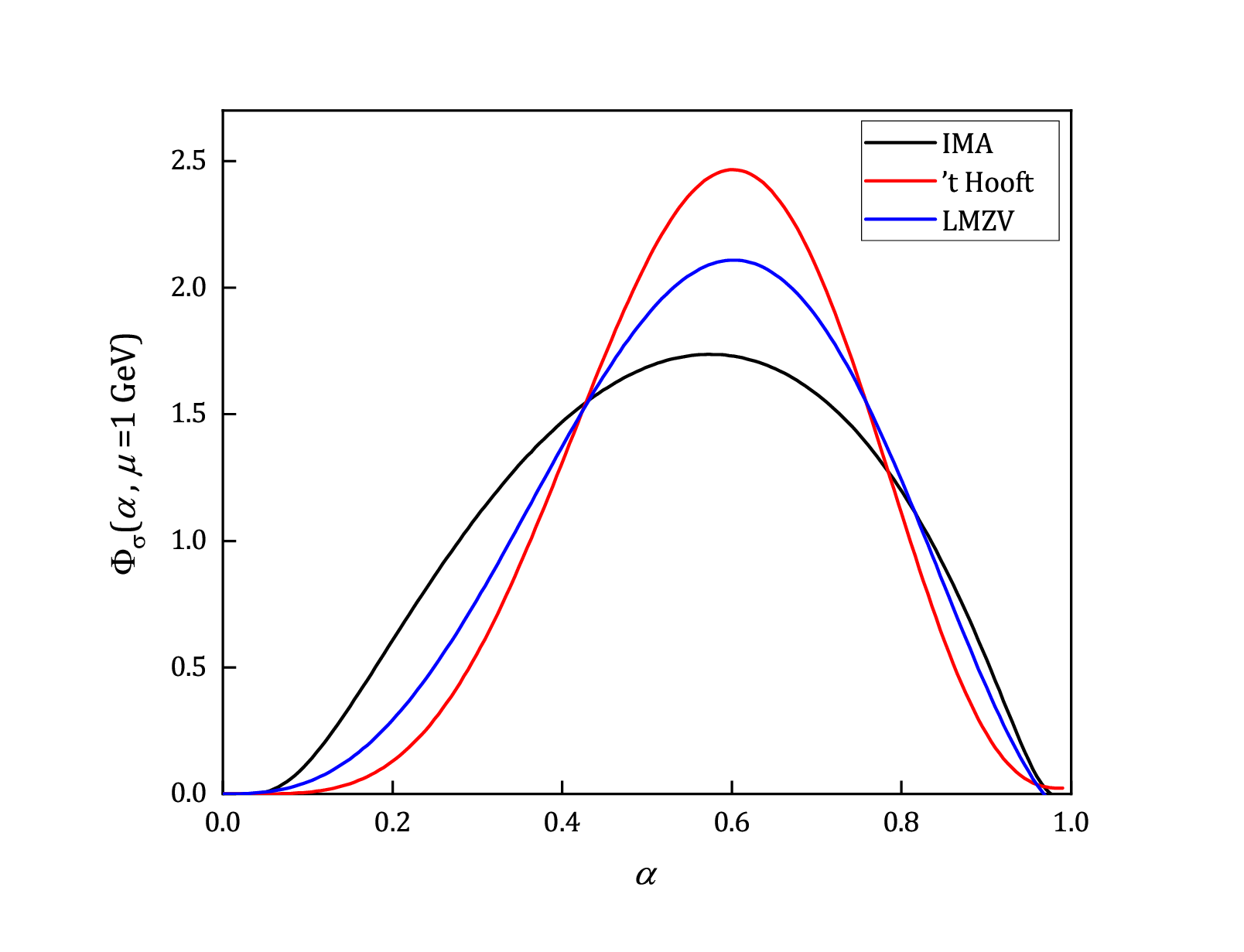}
\caption{Holographic DAs $\Phi$, $\Phi_s$, and $\Phi_\sigma$ of the
$K_0^*(1430)$ meson obtained within the IMA, 't Hooft, and LMZV
frameworks, shown as functions of $\alpha$ at $\mu = 1~\mathrm{GeV}$
and normalized to unity.} \label{F502}
\end{center}
\end{figure}

Figure \ref{F502} displays the holographic DAs of the $K_0^*(1430)$
meson obtained within the IMA, 't Hooft, and LMZV frameworks. The
figure shows the twist-$2$ DA $\Phi$, as well as the scalar and
tensor twist-$3$ DAs, $\Phi_s$ and $\Phi_\sigma$. All distributions
are plotted as functions of the longitudinal momentum fraction
$\alpha$ at the scale $\mu = 1~\mathrm{GeV}$ and are normalized to
unity.

As shown in Fig. \ref{F502}, the peak positions of the DA $\Phi$
exhibit clear quantitative differences among the models. Within the
IMA framework, the distribution reaches its maximum at $\alpha
\simeq 0.3$, whereas in the 't Hooft and LMZV approaches the peak
shifts toward $\alpha \simeq 0.5$ and slightly above. This shift
points to a different pattern of longitudinal momentum sharing, with
a comparatively larger fraction of momentum carried by the strange
quark in the 't Hooft and LMZV cases. By contrast, the IMA result
corresponds to a more moderately distributed momentum profile.
Moreover, the IMA curve displays a broader shape, indicating a
smoother longitudinal distribution. This feature can be attributed
to the absence of an explicit dynamical longitudinal confining
interaction in that framework. In contrast, the close agreement in
both peak position and overall shape between the 't Hooft and LMZV
results reflects their common dynamical treatment of longitudinal
confinement. Despite these quantitative differences, all three
approaches reproduce the characteristic asymmetric structure of the
scalar meson DA, indicating that the overall momentum-sharing
pattern is a robust feature of the light-front description.

The scalar twist-$3$ DA $\Phi_s(\alpha,\mu)$, shown in Fig.
\ref{F502}, develops a maximum near $\alpha \simeq 0.6$. Compared
with the twist-$2$ case, the scalar twist-$3$ amplitudes exhibit
enhanced sensitivity to the modeling of longitudinal dynamics. The
IMA distribution remains broader and flatter, whereas the 't Hooft
and LMZV distributions develop more pronounced peaks. Among these,
the 't Hooft curve reaches slightly higher values, while the LMZV
peak is somewhat lower. The close numerical agreement between the 't
Hooft and LMZV results suggests that the detailed structure of the
scalar twist-$3$ amplitude is governed primarily by the
implementation of longitudinal confinement, while transverse
dynamics plays a secondary role.

The tensor twist-$3$ DA $\Phi_\sigma(\alpha,\mu)$, also displayed in
Fig. \ref{F502}, develops a maximum around $\alpha \simeq 0.6$,
confirming the asymmetric momentum-sharing pattern observed in the
scalar twist-$3$ case. Once again, the 't Hooft prediction exhibits
the highest peak, followed closely by the LMZV result, whereas the
IMA distribution remains broader and lower in magnitude. Although
the overall shapes are qualitatively similar across the three
frameworks, differences in peak height and width are more pronounced
than in the twist-$2$ case. This behavior indicates that the tensor
twist-$3$ amplitude provides a particularly sensitive probe of the
longitudinal confinement mechanism encoded in the light-front wave
functions.

Overall, Fig. \ref{F502} shows that, although the qualitative
features of the $K_0^*(1430)$ DAs are preserved across different
holographic frameworks, their quantitative behavior- especially for
the twist-$3$ components- depends sensitively on the implementation
of longitudinal confinement.

To obtain a clearer and more direct comparison between different
models, the twist-filtered quantity $\delta\Phi(\alpha,\mu)$ is
introduced, which represents the part of the twist-$2$ distribution
amplitude that goes beyond its universal asymptotic form. In this
construction, the full twist-$2$ DA is written as $\Phi(\alpha,\mu)
= \Phi_{\mathrm{asy}}(\alpha,\mu) + \delta\Phi(\alpha,\mu)$, where
$\Phi_{\mathrm{asy}}(\alpha,\mu) = 6\alpha(1-\alpha)$ is the
standard QCD asymptotic shape. The remaining part,
$\delta\Phi(\alpha,\mu)$, contains all non-asymptotic information
and therefore captures the model-dependent nonperturbative effects.
By definition, it satisfies
\begin{equation}\label{eq503}
\int_0^1 d\alpha \, \delta\Phi(\alpha,\mu) = 0,
\end{equation}
so this separation is a standard reparametrization in QCD and does
not introduce any additional physical assumptions.

\begin{figure}[h]
\includegraphics[width=9.75cm,height=8cm]{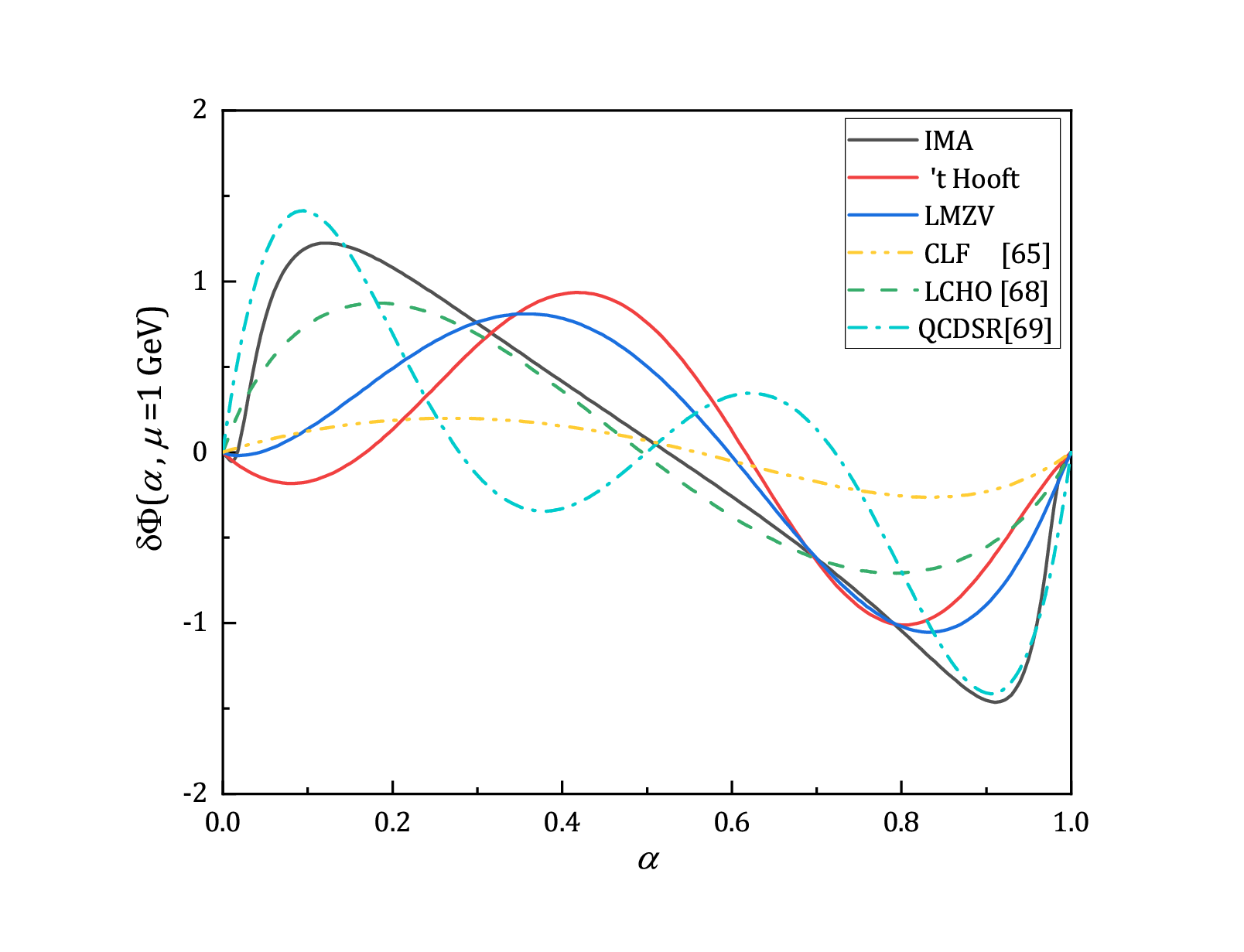}
\caption{Twist-filtered non-asymptotic contribution $\delta\Phi$,
compared with the predictions from CLF \cite{ChZhZha}, LCHO
\cite{DongHuang}, and QCDSR \cite{CheChuYan}.} \label{F503}
\end{figure}

This twist-filtered quantity provides a cleaner way to examine the
differences between theoretical approaches, since it removes the
universal asymptotic contribution and highlights only the
model-specific structures. The resulting twist-filtered distribution
amplitudes are shown in Fig. \ref{F503}, including the predictions
from the IMA, 't Hooft, and LMZV frameworks, as well as the CLF
\cite{ChZhZha}, light-cone harmonic oscillator (LCHO) model
\cite{DongHuang}, and QCDSR \cite{CheChuYan} results.

To eliminate flavor-symmetry breaking arising from unequal quark
masses, equal masses ($m_u = m_s$) are assumed, thereby removing
only explicit mass effects. Nevertheless, even in this limit the
distribution is not expected to be perfectly symmetric around
$\alpha = 0.5$. The residual asymmetry observed in the
twist-filtered amplitudes- particularly in the 't Hooft and LMZV
frameworks- reflects the model-dependent nonperturbative
longitudinal dynamics encoded in the light-front wave functions
rather than originating solely from quark-mass differences.
Consequently, the twist-filtered distributions retain noticeable
asymmetries even in the equal-mass limit, confirming that these
effects originate from nonperturbative longitudinal dynamics rather
than from explicit mass splittings. This observation is fully
consistent with the trends identified in Fig. \ref{F502}.

The $\xi$-moments provide a compact and systematic characterization
of the longitudinal structure of light-front DAs. Defining
\begin{equation}\label{eq504}
\xi = 2\alpha - 1,
\end{equation}
with $\alpha$ denoting the longitudinal momentum fraction carried by
the quark, the $n$-th $\xi$-moment of a DA $\Phi(\alpha,\mu)$ at the
renormalization scale $\mu$ is given by
\begin{equation}\label{eq505}
\langle \xi^n \rangle(\mu) = \int_0^1 d\alpha \, (2\alpha - 1)^n \,
\Phi(\alpha,\mu).
\end{equation}

Even $\xi$-moments probe the overall width and the symmetric core of
the distribution, whereas odd moments are directly sensitive to the
degree of longitudinal asymmetry between quark and antiquark
momentum fractions. Consequently, the hierarchy of $\xi$-moments
effectively isolates the symmetric and antisymmetric contributions
to the DA without requiring an explicit decomposition under $\alpha
\to 1-\alpha$. This structure provides a transparent and
model-independent framework for interpreting the numerical results
summarized in Table \ref{T504}, particularly in assessing the impact
of longitudinal confinement dynamics encoded in the LFWF.
\begin{table}[ht]
\caption{Predicted values for the $\xi$-moments related to the DAs
$\Phi, \Phi_{s}$, and $\Phi_{\sigma}$ of $K^{*}_0(1430)$ at the
scale $\mu = 1$ GeV.} \label{T504}
\begin{ruledtabular}
\begin{tabular}{cccccccc}
$\xi$-moments &IMA  & 't Hooft & LMZV & QCDSR \cite{DongHuang} & QCDSR \cite{CheChuYan} & QCDSR \cite{HanWuFu} & QCDSR \cite{CDLu}\\
\hline
$\langle \xi^1\rangle $        & $-0.66^{+0.31}_{-0.43}$ & $-0.08^{+0.07}_{-0.08}$ & $-0.19^{+0.11}_{-0.13}$ & $-0.261_{-0.071}^{+0.056}$ & $-0.35^{+0.08}_{-0.08}$                          \\
$\langle \xi^2\rangle$         & $+0.02^{+0.02}_{-0.02}$ & $+0.03^{+0.02}_{-0.03}$ & $+0.03^{+0.02}_{-0.04}$ & $+0.007_{-0.006}^{+0.005}$ &                                                              \\
$\langle \xi^3\rangle $        & $-0.30^{+0.13}_{-0.18}$ & $-0.04^{+0.02}_{-0.03}$ & $-0.09^{+0.04}_{-0.08}$ & $-0.177_{-0.045}^{+0.034}$ & $-0.23^{+0.06}_{-0.06}$                                  \\
$\langle \xi^1_{s}\rangle$     & $+0.10^{+0.02}_{-0.02}$ & $+0.16^{+0.01}_{-0.01}$ & $+0.16^{+0.01}_{-0.01}$ &                            &                         & $+0.003\sim0.004$                         &$+0.006\sim0.014$                      \\
$\langle \xi^2_{s}\rangle$     & $+0.27^{+0.02}_{-0.02}$ & $+0.14^{+0.01}_{-0.01}$ & $+0.20^{+0.02}_{-0.02}$ &                            &                         & $+0.352\sim0.358$                         &$+0.290\sim0.330$                      \\
$\langle \xi^1_{\sigma}\rangle$& $+0.09^{+0.04}_{-0.05}$ & $+0.16^{+0.06}_{-0.08}$ & $+0.14^{+0.06}_{-0.08}$ &                            &                         & $+0.010\sim0.030$                         &$+0.022\sim0.033$                      \\
$\langle \xi^2_{\sigma}\rangle$& $+0.16^{+0.03}_{-0.03}$ & $+0.12^{+0.04}_{-0.08}$ & $+0.14^{+0.05}_{-0.06}$ &                            &                         & $+0.206\sim0.208$                         &$+0.200\sim0.250$                      \\
\end{tabular}
\end{ruledtabular}
\end{table}
As shown in the tables, the second moment $\langle \xi^2 \rangle$ of
the leading-twist DA remains small and positive in all three
frameworks, indicating a comparable overall width and a largely
model-independent symmetric core. In contrast, a pronounced
hierarchy emerges in the odd moments. The IMA framework predicts
significantly larger magnitudes for $\langle \xi \rangle$ and
$\langle \xi^3 \rangle$ than both the 't Hooft and LMZV models,
signalling a substantially stronger longitudinal asymmetry in the
leading-twist DA. The latter two approaches yield numerically close
results, reflecting similar longitudinal wave-function profiles and
closely related implementations of confinement dynamics.

A qualitatively similar pattern is observed in the twist-$3$ sector.
The odd moments of $\Phi_s$ and $\Phi_\sigma$ display moderate model
dependence, while the corresponding even moments are positive and of
comparable size across all approaches. This confirms the relative
stability of the symmetric core of the twist-$3$ DAs, with model
sensitivity again residing primarily in the antisymmetric component.

Overall, the analysis demonstrates that the dominant model
dependence is concentrated in the antisymmetric part of the
twist-$2$ DA, which directly probes longitudinal momentum asymmetry.
Even moments and twist-$3$ observables exhibit significantly milder
variations, indicating a robust symmetric light-front structure. The
observed differences among models can be traced to their distinct
realizations of confinement: the IMA framework, characterized by
exponential endpoint suppression, yields broader longitudinal
profiles, whereas the LMZV approach, with its power-law behavior,
produces comparatively narrower distributions. Compared to QCDSR
predictions, the holographic distribution amplitudes exhibit
smoother endpoint behavior, a feature that can be attributed to the
confinement mechanism intrinsic to the light-front holographic
framework.

\begin{figure}[h]
\includegraphics[width=8.75cm,height=8cm]{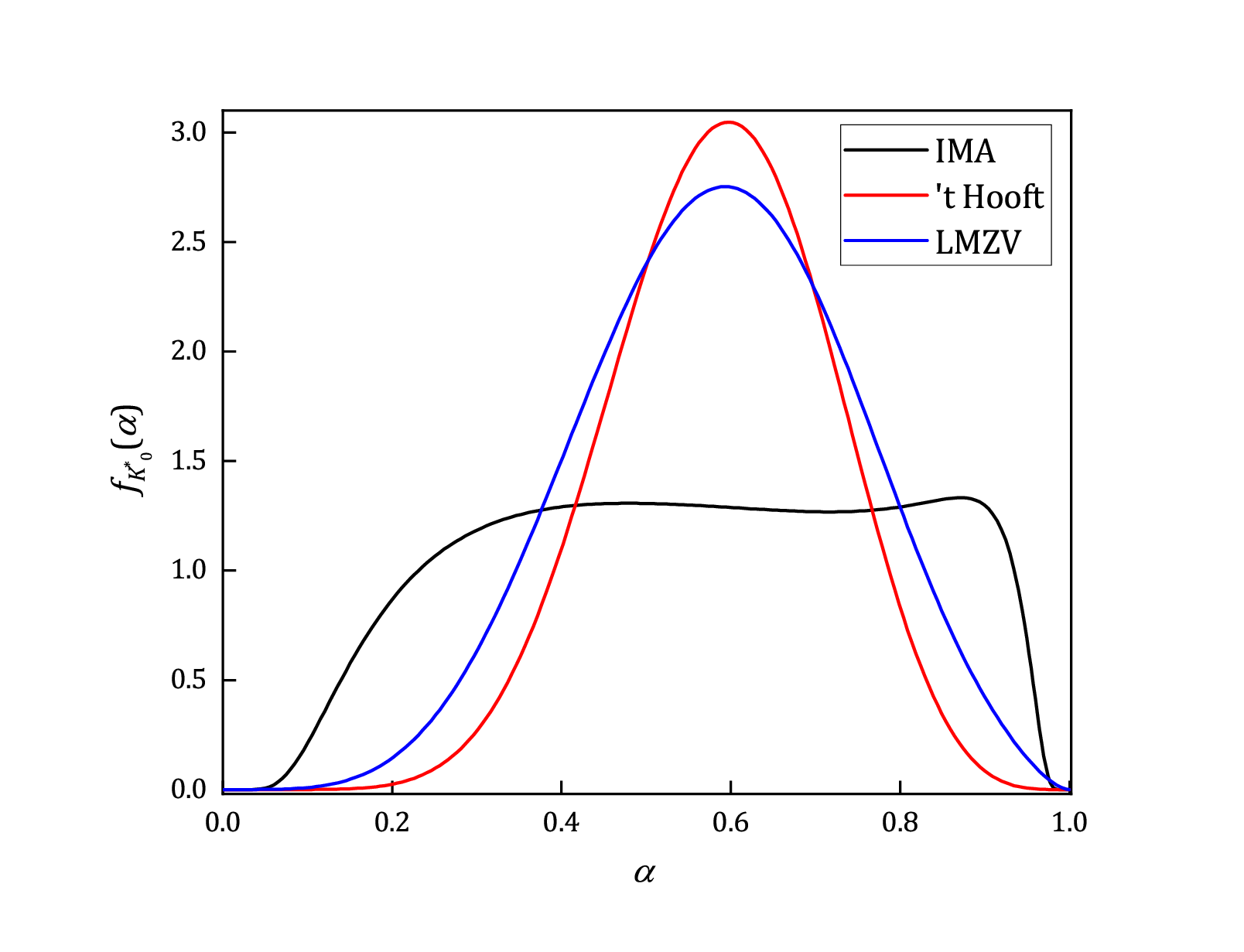}
\caption{PDF of the $K_{0}^{*}(1430)$ meson as a function of the
longitudinal momentum fraction $\alpha$.} \label{F505}
\end{figure}

The holographic valence PDF of the $K_0^*(1430)$ meson is defined as
\cite{BrTe,AhmMacKKa}
\begin{align}
f_{K_0^*}(\alpha) = \sum_{h, \bar{h}} \int d^2 \mathbf{r_\perp} \,
|\Psi_{h, \bar{h}}(\mathbf{r_\perp}, \alpha)|^2. \label{eq512}
\end{align}
This observable encodes the longitudinal momentum distribution of
the valence constituents in the scalar meson. Within the light-front
holographic framework, it is derived from the same underlying wave
function that determines the corresponding distribution amplitude,
thereby ensuring consistency between spatial and momentum-space
descriptions. The resulting valence PDF $f_{K_0^*}(\alpha)$ is
displayed in Fig. \ref{F505}.

All three approaches exhibit a pronounced asymmetry with the maximum
shifted toward $\alpha > 0.5$, a direct manifestation of $SU(3)$
flavor-symmetry breaking. Because the strange quark is heavier than
the light antiquark, it carries a larger share of the longitudinal
momentum, leading to a systematic distortion of the distribution
away from the symmetric point. Beyond this universal trend, the
detailed shapes of the PDFs reveal important dynamical differences
among the three models. The IMA framework generates a comparatively
broad and nearly flat distribution, with a mild suppression only
near the endpoints. Such a profile indicates that the wave function
in this model supports a wide range of longitudinal momentum
configurations, including substantially asymmetric ones. In
contrast, the 't Hooft model produces a sharply peaked and
significantly narrower distribution. The enhanced localization
around the peak reflects the stronger impact of longitudinal
confinement in this approach, which effectively suppresses
configurations with large momentum imbalance. The rapid fall-off
toward both endpoints is consistent with the underlying structure of
the 't Hooft kernel and its characteristic treatment of chiral and
confinement dynamics in $1+1$ dimensions. The LMZV distribution lies
between these two extremes. It is narrower than the IMA result but
broader than the 't Hooft one, and its peak height correspondingly
takes an intermediate value. This behavior signals that the
effective holographic potential in the LMZV framework induces a
nontrivial but moderate degree of localization in longitudinal
momentum space. The controlled suppression of the endpoints further
reflects the interplay between longitudinal constraints and the
modified holographic embedding of this model.

Overall, the comparison demonstrates how the underlying confining
mechanisms and mass-dependent dynamics of each framework shape the
longitudinal momentum structure of the scalar meson. The distinct
widths, peak positions, and endpoint behaviors encapsulate the
differences in how each model implements effective QCD dynamics, and
thus they provide a clear and physically transparent basis for
evaluating their internal consistency and phenomenological
implications.

After establishing the DAs of the $K_0^*(1430)$ meson, we proceed to
the transition form factors relevant for the rare semileptonic decay
$B \to K_{0}^{*}\, \ell^{+} \ell^{-}$, with $\ell=\mu$ or $\tau$.
Imposing Lorentz invariance and parity conservation, the
corresponding hadronic matrix elements can be parameterized as
\cite{WanAsl}
\begin{align}
\langle K_{0}^{*}(p)|\bar{s}\gamma_{\mu}\gamma_{5} b |B(p+q)\rangle
&= -i\left[f_{+}(q^{2})p_{\mu}+f_{-}(q^{2})q_{\mu}\right],
\label{eq513}\\
\langle K_{0}^{*}(p)|\bar{s}\sigma_{\mu\nu}\gamma_{5}q^{\nu}b
|B(p+q)\rangle &= -\frac{f_{T}(q^{2})}{m_{B}+m_{K_{0}^{*}}}
\left[(2p+q)_{\mu}q^{2}-(m_{B}^{2}-m_{K_{0}^{*}}^{2})q_{\mu}\right].
\label{eq514}
\end{align}
Within the LCSR framework, the form factors $f_{+}(q^{2})$,
$f_{-}(q^{2})$, and $f_{T}(q^{2})$ are expressed in terms of the
$K_0^*$ DAs $\Phi$, $\Phi_s$, and $\Phi_{\sigma}$, as given in Eqs.
(\ref{eq515})-(\ref{eq517}) \cite{WanAsl}:
\begin{align}
f_{+}(q^{2}) &= \frac{(m_{b}+m_{s})\,
\bar{f}_{_{K_0^*}}}{2\,m_{B}^{2}f_{B}}\, e^{\frac{m_{B}^{2}}{M^{2}}}
\Bigg\{ \int_{\alpha_{0}}^{1}\frac{d\alpha}{\alpha}
e^{-\frac{m_{b}^{2}+\alpha(1-\alpha)p^{2}-(1-\alpha)q^{2}} {\alpha
M^{2}}} \Bigg[ -m_{b}\Phi(\alpha) +\alpha m_{K_0^*}\Phi_{s}(\alpha)
+\frac{m_{K_0^*}}{3}\Phi_{\sigma}(\alpha)
\nonumber\\
&\hspace{3cm} +\frac{m_{K_0^*}}{6\alpha M^{2}}
\left(m_{b}^{2}+\alpha^{2}p^{2}+q^{2}\right) \Phi_{\sigma}(\alpha)
\Bigg] +\frac{m_{K_0^*}}{6} \frac{m_{b}^{2}-\alpha_{0}p^{2}+q^{2}}
{m_{b}^{2}+\alpha_{0}^{2}p^{2}-q^{2}} \Phi_{\sigma}(\alpha_{0})
e^{-s_{0}/M^{2}} \Bigg\},
\label{eq515}\\
f_{-}(q^{2}) &= \frac{(m_{b}+m_{s})\,
\bar{f}_{_{K_0^*}}}{m_{B}^{2}f_{B}}\, e^{\frac{m_{B}^{2}}{M^{2}}}
\Bigg\{ \int_{\alpha_{0}}^{1}\frac{d\alpha}{\alpha}
e^{-\frac{m_{b}^{2}+\alpha(1-\alpha)p^{2}-(1-\alpha)q^{2}} {\alpha
M^{2}}} \Bigg[ m_{K_0^*}\Phi_{s}(\alpha)
+\frac{m_{K_0^*}}{6\alpha}\Phi_{\sigma}(\alpha)
\nonumber\\
&\hspace{3cm} -\frac{m_{K_0^*}}{6\alpha^{2}M^{2}}
\left(m_{b}^{2}+\alpha^{2}p^{2}-q^{2}\right) \Phi_{\sigma}(\alpha)
\Bigg] -\frac{m_{K_0^*}}{6\alpha_{0}} \Phi_{\sigma}(\alpha_{0})
e^{-s_{0}/M^{2}} \Bigg\},
\label{eq516}\\
f_{T}(q^{2}) &= \frac{(m_{b}+m_{s})(m_{B}+m_{K_0^*})\,
\bar{f}_{_{K_0^*}}}{m_{B}^{2}f_{B}}\, e^{\frac{m_{B}^{2}}{M^{2}}}
\Bigg\{ -\frac{1}{2} \int_{\alpha_{0}}^{1}\frac{d\alpha}{\alpha}
e^{-\frac{m_{b}^{2}+\alpha(1-\alpha)p^{2}-(1-\alpha)q^{2}} {\alpha
M^{2}}} \Bigg[ \Phi(\alpha) -\frac{m_{B}m_{K_0^*}}{3\alpha M^{2}}
\Phi_{\sigma}(\alpha) \Bigg]
\nonumber\\
&\hspace{3cm} +\frac{m_{b}m_{K_0^*}}{6}
\frac{\Phi_{\sigma}(\alpha_{0})}
{m_{b}^{2}+\alpha_{0}^{2}p^{2}-q^{2}} e^{-s_{0}/M^{2}} \Bigg\}.
\label{eq517}
\end{align}
In these expressions, $q^{2}$ denotes the squared momentum transfer,
and the lower integration limit $\alpha_{0}$ is determined by the
continuum threshold according to
\begin{equation}\label{eq518}
\alpha_{0}= \frac{\sqrt{(s_{0}-q^{2}-p^{2})^{2}
+4p^{2}(m_{b}^{2}-q^{2})} -(s_{0}-q^{2}-p^{2})}{2p^{2}},
\end{equation}
where $p^{2}=m_{K_0^*}^{2}$. The Borel parameter $M^{2}$ and the
continuum threshold $s_{0}$ are auxiliary quantities introduced in
the sum rule procedure. Consequently, the physical form factors are
expected to exhibit only a mild dependence on these parameters
within the corresponding stability window.

\begin{figure}[h]
\includegraphics[width=5.5cm,height=5cm]{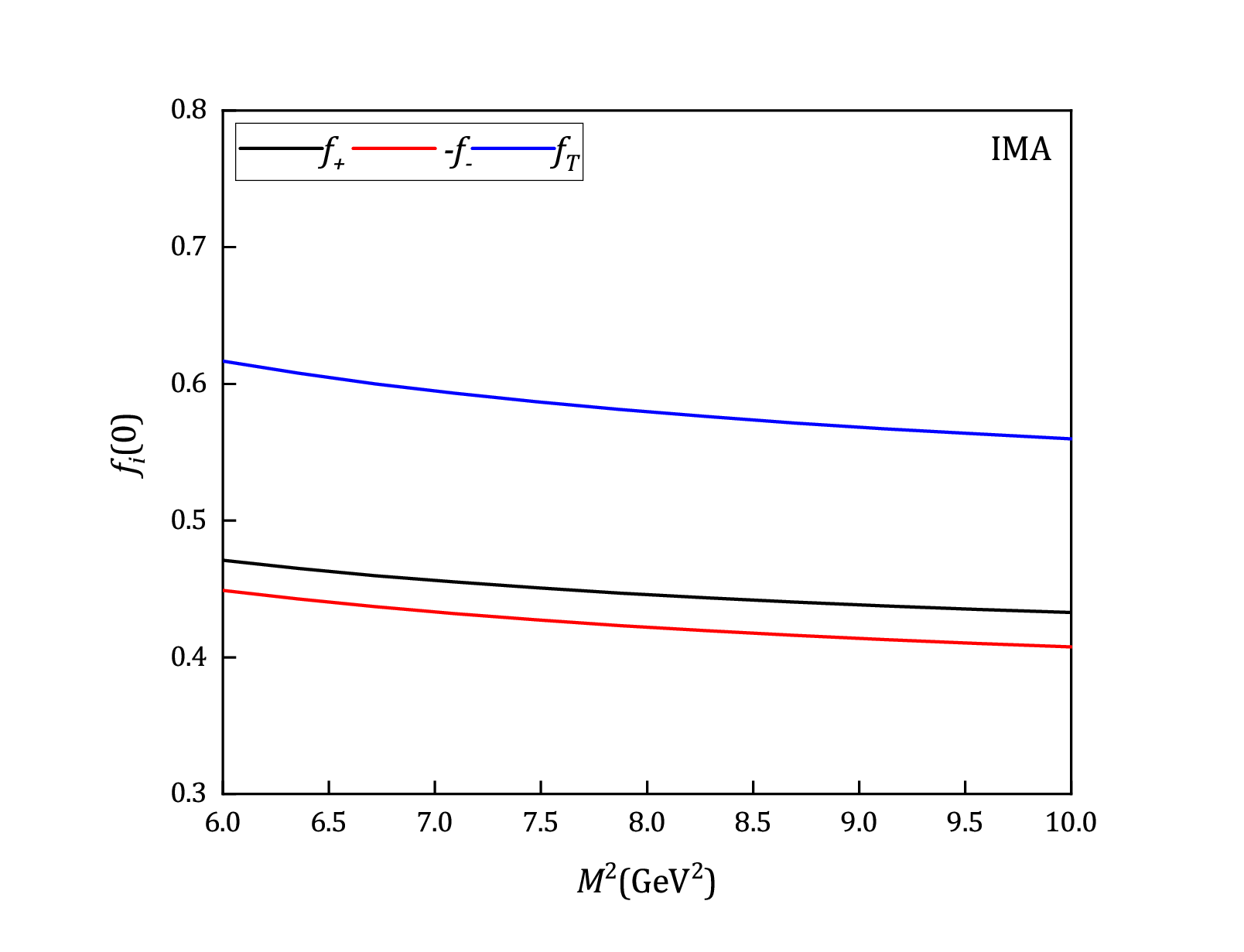}
\includegraphics[width=5.5cm,height=5cm]{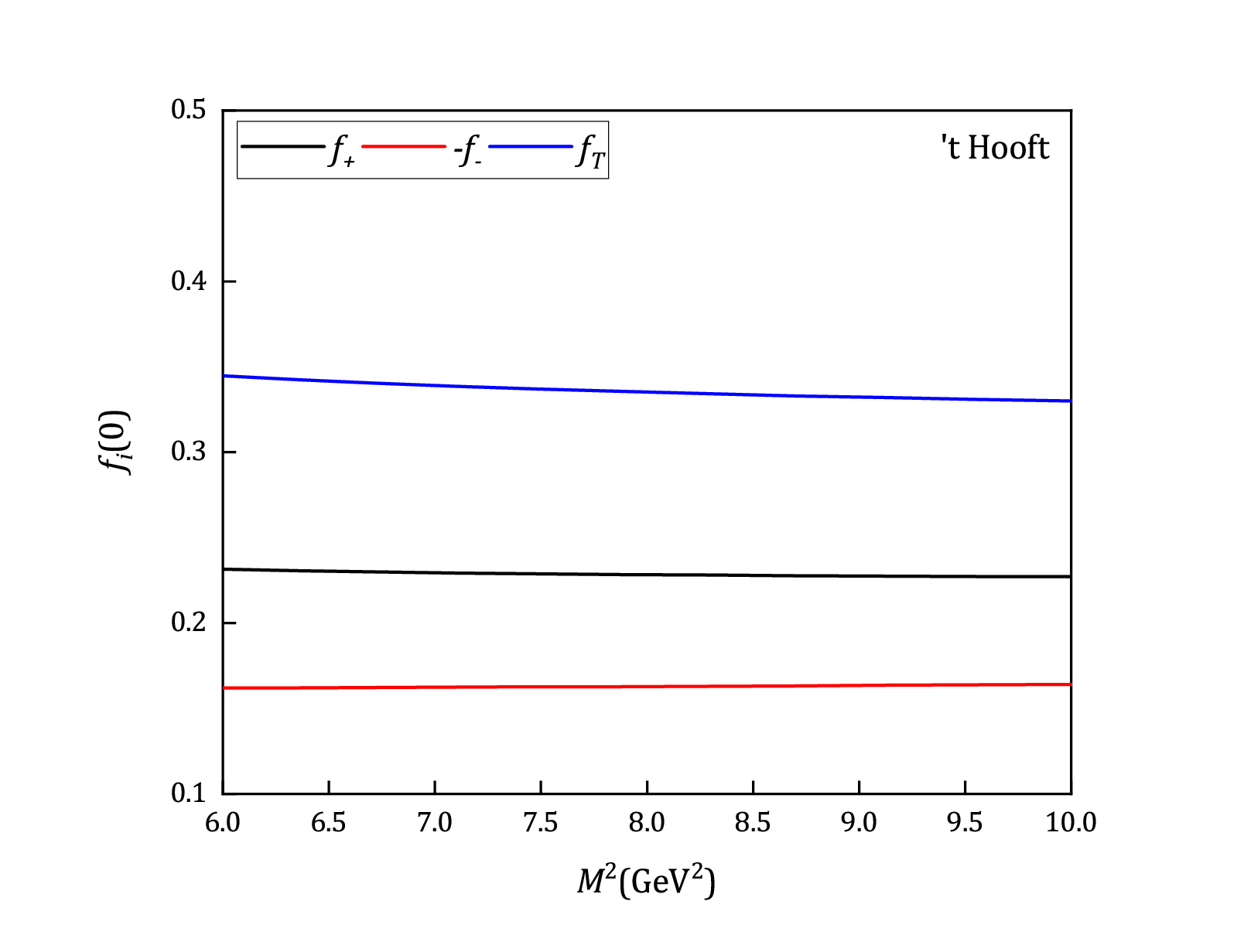}
\includegraphics[width=5.5cm,height=5cm]{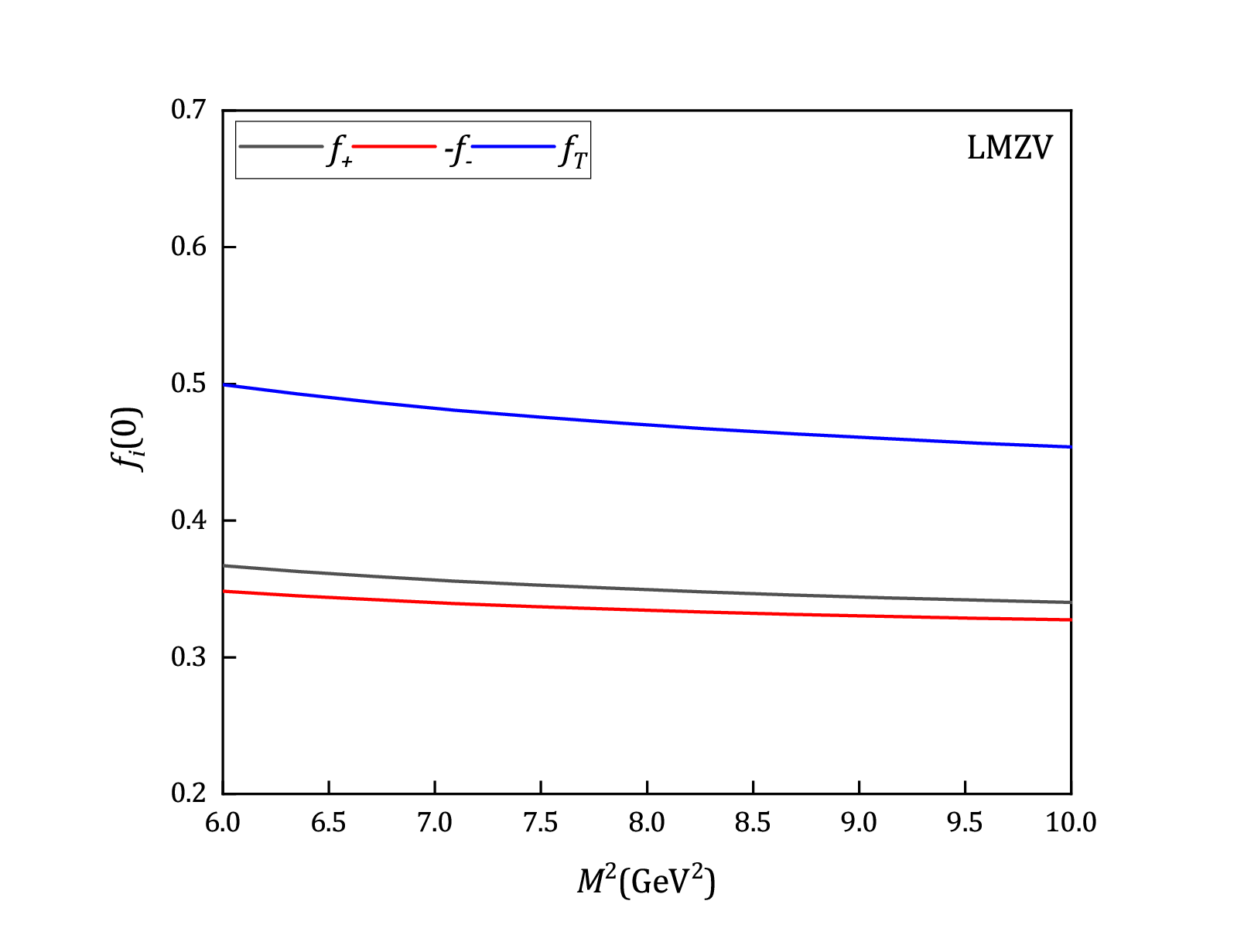}
\caption{Dependence of the form factors $f_{+}$, $f_{-}$, and
$f_{T}$ on the Borel parameter $M^{2}$ at $q^2=0$, obtained using
the $K_0^*$ DAs in the IMA, 't Hooft, and LMZV approaches.}
\label{F506}
\end{figure}

The continuum threshold is taken as $s_{0}=(33 \pm
1)\,\mathrm{GeV}^{2}$ \cite{SunLi,ZGWang}, which is consistent with
previous LCSR analyses of the scalar $B$ meson channel. Using the
$K_0^*$ DAs via the IMA, 't Hooft, and LMZV and also decay constant
values of the $K_{0}^{*}$ meson obtained within the LFHQCD framework
in Table \ref{T503}, together with the input parameters $m_{b}=(4.81
\pm 0.03)$ GeV, $m_{B}=5.27$ GeV \cite{PDG}, and $f_{B}=(190 \pm
10)$ MeV \cite{AokAokBec}, and setting $q^{2}=0$ in Eqs.
(\ref{eq515})-(\ref{eq517}), a good stability of the form factors
with respect to the Borel parameter is observed in the interval
$6\,\mathrm{GeV}^{2}\le M^{2}\le10\,\mathrm{GeV}^{2}$. The
dependence of the form factors $f_{+}$, $f_{-}$, and $f_{T}$ on the
Borel parameter $M^{2}$ is illustrated in Fig. \ref{F506}.
For the subsequent numerical analysis, we adopt $M^{2}=8 \pm
2\,\mathrm{GeV}^{2}$.

The resulting values of $f_+(0)$, $f_-(0)$, and $f_T(0)$, together
with the uncertainties arising from the input parameters, are listed
in Table \ref{T506}. For comparison, results from other theoretical
approaches, including LCSR
\cite{HanWuFu,WanAsl,SunLi,Khosravi1,ZGWang}, QCDSR \cite{AliAzi},
PQCD \cite{RunHui}, LFQM \cite{ChGeLi}, CLF \cite{CheChuHwa}, and
MSSM \cite{AsluWang}, are also presented.
\begin{table}[ht]
\caption{Form factor values for the semileptonic decay $B \to
K_0^*(1430)\ell^+\ell^-$ at $q^2=0$ from various theoretical
approaches. All uncertainties are taken from the original
references. } \label{T506}
\begin{ruledtabular}
\begin{tabular}{lcccc}
Method                    & Ref
& $f_+(0)$
& $f_-(0)$
& $f_T(0)$ \\
\hline

IMA               & Ours
&$0.46^{+0.29}_{-0.25}$
&$-0.43^{+0.25}_{-0.28}$
&$0.60^{+0.38}_{-0.35}$ \\

't Hooft                 & Ours
&$0.23^{+0.10}_{-0.11}$
&$-0.16^{+0.13}_{-0.14}$
&$0.34^{+0.20}_{-0.22}$ \\

LMZV                     & Ours
&$0.36^{+0.15}_{-0.14}$
&$-0.33^{+0.19}_{-0.21}$
&$0.49^{+0.20}_{-0.34}$ \\

LCSR                     & \cite{Khosravi1}
&$0.53^{+0.28}_{-0.22}$
&$-0.51^{+0.13}_{-0.36}$
&$0.72^{+0.39}_{-0.31}$ \\

LCSR                     & \cite{ZGWang}
&$0.52^{+0.07}_{-0.07}$
&$-0.27^{+0.06}_{-0.06}$
&$0.66^{+0.10}_{-0.10}$ \\

LCSR                     & \cite{HanWuFu}
&$0.45$
&$-0.28$
&$0.46$ \\

LCSR (S1)                & \cite{SunLi}
&$0.17$
&$-0.17$
&$0.24$ \\

LCSR (S2)                & \cite{SunLi}
&$0.49$
&$-0.49$
&$0.69$ \\

LCSR                     & \cite{WanAsl}
&$0.49$
&$-0.41$
&$0.60$ \\

QCDSR                    & \cite{AliAzi}
&$0.31$
&$-0.31$
&$-0.26$ \\

PQCD (S2)                & \cite{RunHui}
&$0.60$
&$-0.60$
&$0.78$ \\

PQCD (S1)                & \cite{RunHui}
&$-0.34$
&
&$-0.44$ \\

LFQM                     & \cite{ChGeLi}
&$-0.26$
&$0.21$
&$-0.34$ \\

CLF                      & \cite{CheChuHwa}
&$0.26$
&
&       \\

MSSM                     & \cite{AsluWang}
&$0.49$
&$-0.41$
&$0.60$ \\
\end{tabular}
\end{ruledtabular}
\end{table}

A comparison of the results listed in Table \ref{T506} reveals a
systematic pattern among the holographic predictions. The IMA
framework produces the largest central values for the form factors,
whereas the 't Hooft model leads to more suppressed results. The
LMZV predictions lies between these two limits. Since all three
calculations are performed under identical numerical conditions, the
observed spread mainly reflects differences in the dynamical
implementation of longitudinal confinement rather than variations in
the input parameters.

The values of $f_+(0)$ reported in the literature cover a relatively
wide range, approximately $0.17$--$0.60$, which reflects the
significant theoretical uncertainty associated with modeling the
internal structure of the scalar meson $K_0^*(1430)$. In particular,
PQCD and some LCSR analyses tend to predict larger magnitudes,
whereas the holographic approaches considered here favor more
moderate values.

The values of the form factors at $q^2=0$ play an important role
since they provide boundary conditions for the description of their
$q^2$ dependence across the physical region. Because differential
decay rates and angular observables depend nonlinearly on these form
factors, even moderate differences in their values at low $q^2$ may
lead to noticeable variations in predicted branching ratios or
asymmetries. Consequently, improved experimental measurements of
these observables could provide useful tests of the underlying
dynamical assumptions of the theoretical models.

In the present analysis, the $K_0^*(1430)$ meson is treated as a
conventional $\bar q q$ scalar state, consistent with the
assumptions adopted in most LCSR, QCDSR, and PQCD studies. Possible
multiquark or molecular components are therefore not considered
within the current framework.

\begin{figure}[h]
\includegraphics[width=5.75cm,height=5cm]{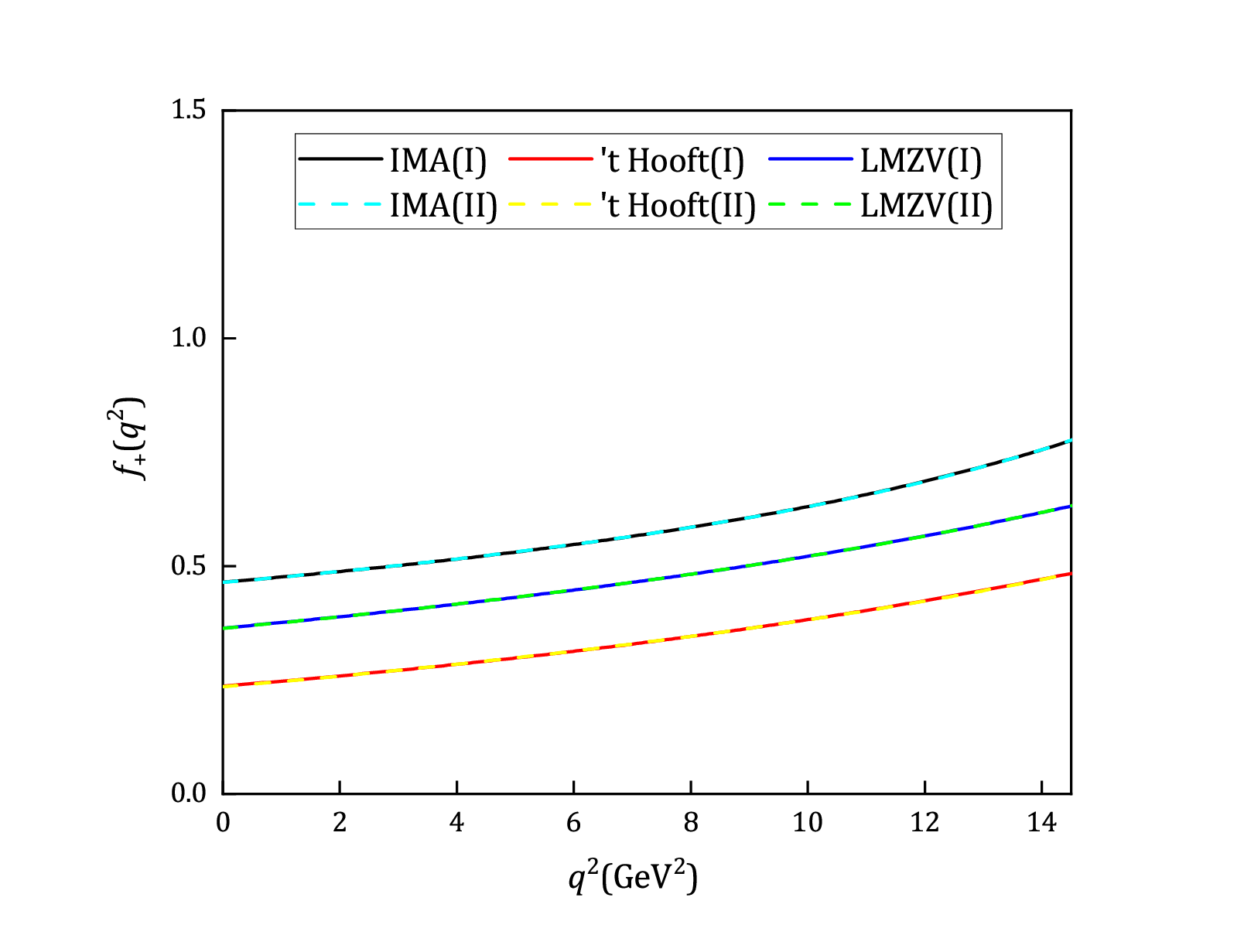}
\includegraphics[width=5.75cm,height=5cm]{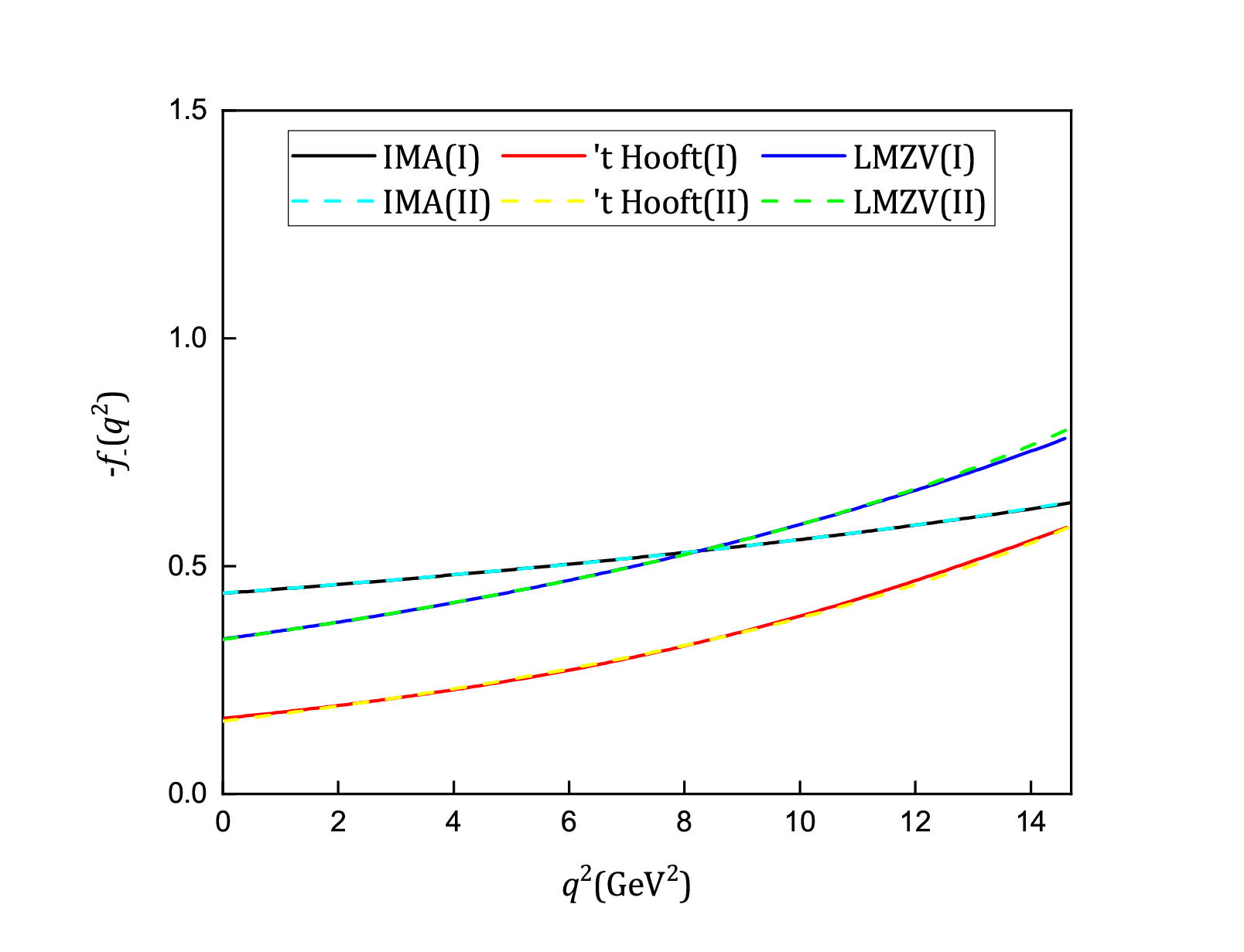}
\includegraphics[width=5.75cm,height=5cm]{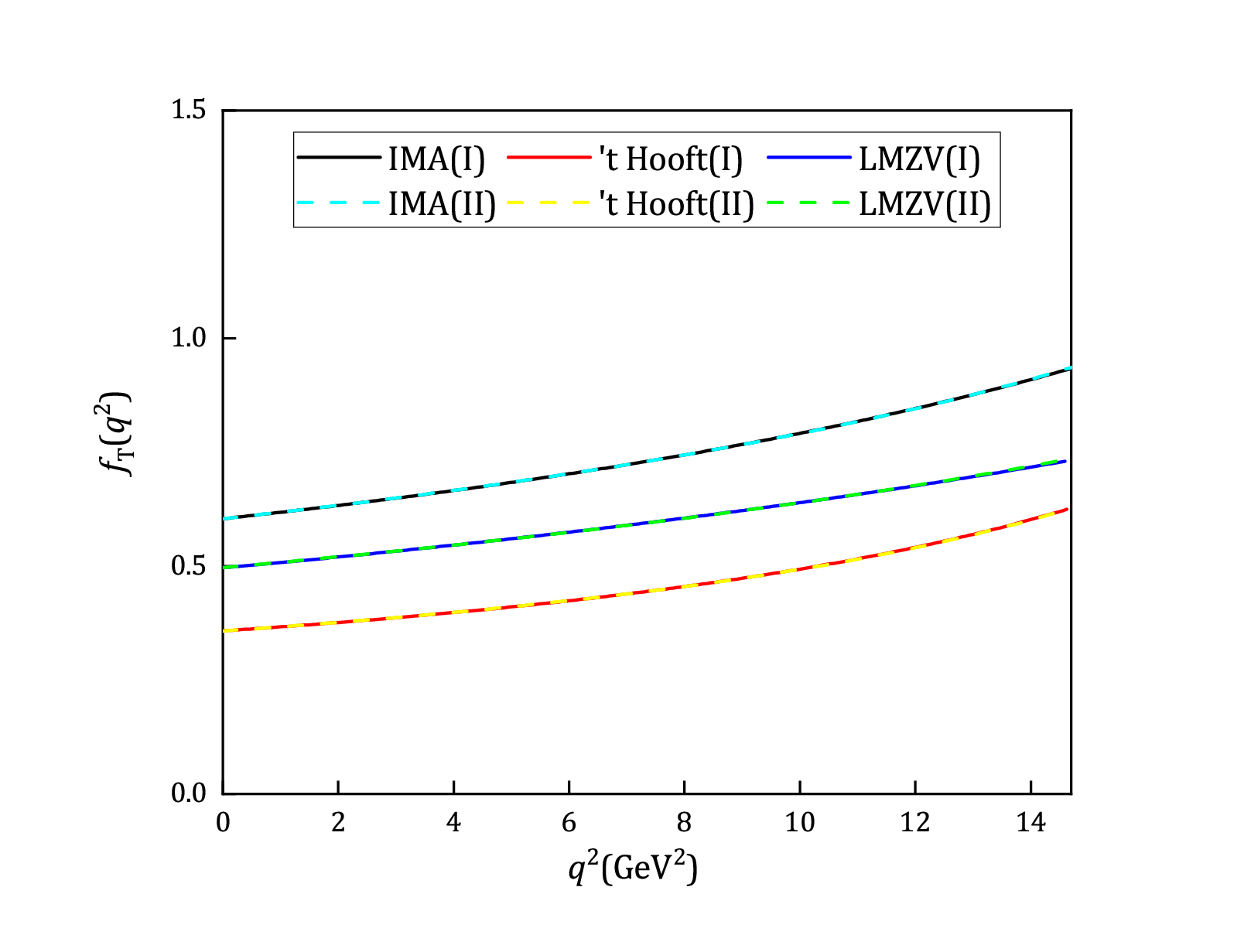}
\caption{The fitted form factors $f_+$, $-f_-$, and $f_T$ for the $B
\to K_0^*$ transition, obtained using the fit functions
$f^{\mathrm{I}}(q^2)$ and $f^{\mathrm{II}}(q^2)$, are shown as
functions of $q^2$ in the IMA, 't Hooft, and LMZV
models.}\label{F507}
\end{figure}

Due to the cutoff inherent in QCD-based approaches, the form factors
can be computed with controlled accuracy only in the
low-to-intermediate kinematic region, $0 \le q^2 \lesssim (m_B -
m_{K_0^*})^2/2$. To extrapolate the results to the full physical
range, $0 \le q^2 \le (m_B-m_{K_0^*})^2$, we employ two commonly
used parametrizations:
\begin{align}
f^{\mathrm{I}}(q^2) &= \frac{f(0)}{1 - \alpha\, s + \beta\, s^2}, \label{eq520}\\
f^{\mathrm{II}}(q^2) &= \frac{1}{1 - s} \sum_{k=0}^{2} b_k \big[
z(q^2) - z(0) \big]^k, \label{eq521}
\end{align}
where $s = q^2/m_B^2$, and
\begin{equation}\label{eq522}
z(q^2) = \frac{\sqrt{t_+ - q^2} - \sqrt{t_+ - t_0}}
     {\sqrt{t_+ - q^2} + \sqrt{t_+ - t_0}},
\end{equation}
with
\begin{equation}\label{eq523}
t_0 = t_+ \Big(1 - \sqrt{1 - \frac{t_-}{t_+}}\Big), \qquad t_{\pm} =
(m_{B} \pm m_{K_0^*})^2 .
\end{equation}
The parameters $(f(0), \alpha, \beta)$ and $(b_0, b_1, b_2)$ are
determined from the fitting procedure. The fitted values for the
three wave function-based models (IMA, 't Hooft, and LMZV) are
collected in Table \ref{T507}.
\begin{table}[ht]
\caption{Values of parameters $(f(0),  \alpha,  \beta)$ and ($b_{0},
b_{1},  b_{2}$) connected to the fit functions $f^{\mathrm{I}}(q^2)$
and $f^{\mathrm{II}}(q^2)$ respectively, for the fitted form factors
of $B \to K_0^*$ transitions of the three models IMA, 't Hooft, and
LMZV.} \label{T507}
\begin{ruledtabular}
\begin{tabular}{cc|ccc|ccc}
Method & Form Factor & $f(0)$ & $\alpha$ & $\beta$ & $b_{0}$ & $b_{1}$ & $b_{2}$ \\
\hline{IMA}
 & $f_{+}$ & $0.46$ & $0.72$ & $-0.07$ & $0.46$ & $0.94$ & $0.32$ \\
 & $f_{-}$ & $-0.43$ & $0.64$  & $0.11$ & $-0.44$ & $ -1.14$ & $1.73$ \\
 & $f_{T}$ & $0.60$ & $ 0.69$ & $0.05$ & $0.60$ & $1.32$ & $-2.25$ \\
\hline{'t Hooft}
 & $f_{+}$ & $0.23$ & $1.37$ & $0.77$ & $0.23$ & $-0.50$ & $-5.63$ \\
 & $f_{-}$ & $-0.16$ & $2.15$  & $ 1.52$ & $-0.15$ & $1.76$ & $5.89$ \\
 & $f_{T}$ & $0.34$ & $0.89$ & $0.20$ & $0.35$ & $0.74$ & $1.11$ \\
\hline{LMZV}
 & $f_{+}$ & $0.36$ & $0.98$ & $0.33$ & $0.36$ & $0.19$ & $-4.52$ \\
 & $f_{-}$ & $-0.33$ & $1.41$  & $0.64$ & $-0.33$ & $1.02$ & $6.32$ \\
 & $f_{T}$ & $0.49$ & $0.74$ & $0.27$ & $0.49$ & $1.18$ & $-3.46$ \\
\end{tabular}
\end{ruledtabular}
\end{table}
The resulting $q^2$ dependence of the fitted form factors $f_i$
$(i=+, -, T)$ is displayed in Fig. \ref{F507}. For each model, both
parametrizations $f^{\mathrm{I}}(q^2)$ and $f^{\mathrm{II}}(q^2)$
are shown.
The two parametrizations lead to numerically very similar curves
over the entire physical range. Their near overlap indicates that
the extrapolation uncertainty associated with the choice of fit
ansatz is small compared with the intrinsic theoretical errors.
Accordingly, in the subsequent phenomenological analysis, we adopt
the results obtained from the simpler form $f^{\mathrm{I}}(q^2)$.

\begin{figure}[h]
\includegraphics[width=9.75cm,height=8cm]{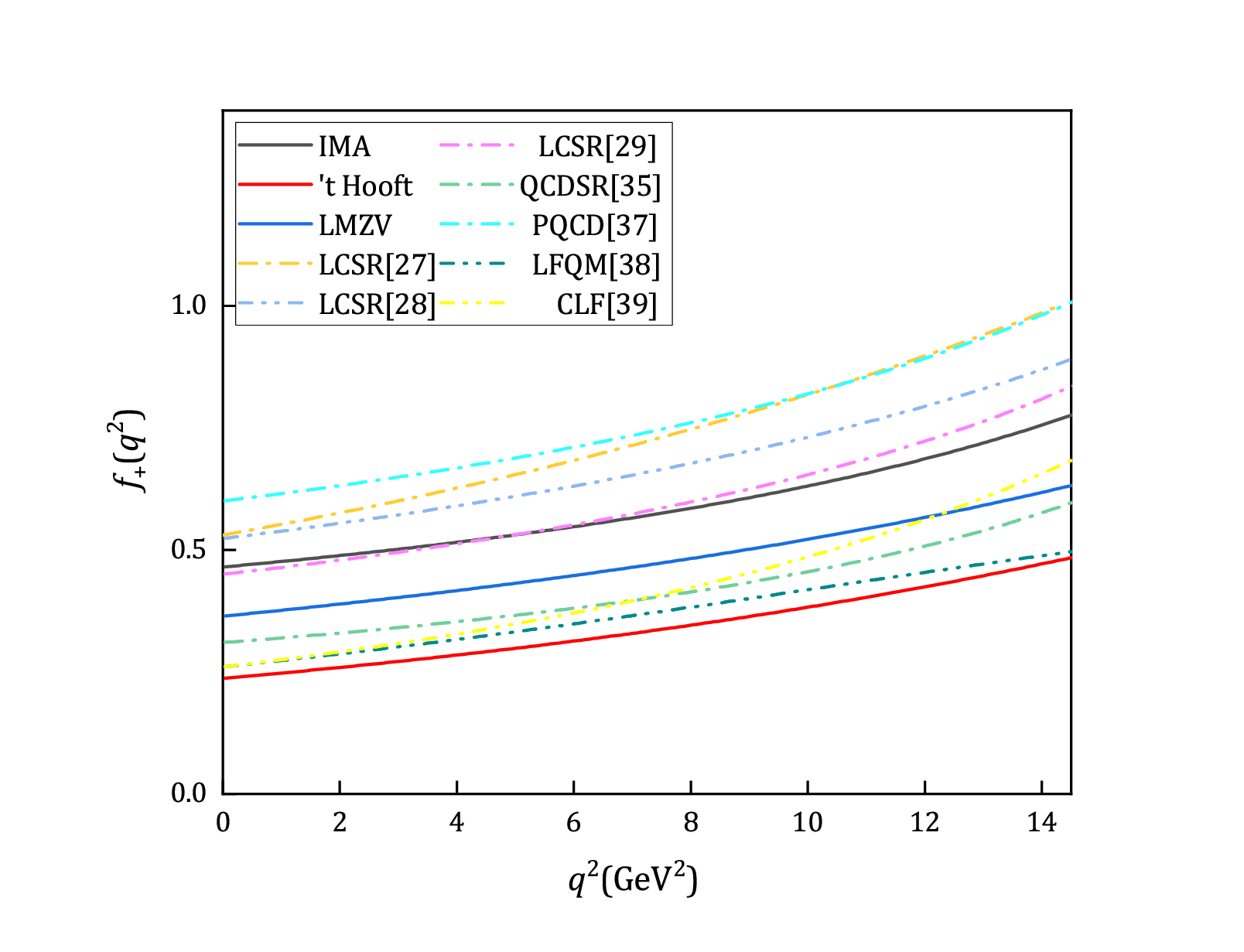}
\caption{Transition form factor $f_{+}(q^2)$ for the $B \to K_0^*\,
\ell^+ \ell^-$ decay as a function of $q^2$ obtained in different
theoretical approaches.} \label{F508}
\end{figure}

Figure \ref{F508} presents a comparison of the transition form
factor $f_{+}(q^{2})$ obtained within our three holographic
frameworks, namely IMA, 't Hooft, and LMZV, together with several
existing determinations in the literature. In particular, our
predictions are confronted with results derived from LCSR
\cite{Khosravi1,ZGWang,HanWuFu}, QCDSR \cite{AliAzi}, PQCD
\cite{RunHui}, LFQM \cite{ChGeLi}, and the CLF \cite{CheChuHwa}
model.
As illustrated in Fig. \ref{F508}, all theoretical approaches
display a monotonically increasing behavior with respect to $q^{2}$,
which is the expected trend resulting from the enhanced overlap
between the initial and final hadronic states at larger momentum
transfer. Despite this common qualitative pattern, the predictions
exhibit sizeable quantitative differences, particularly in the
normalization at $q^{2}=0$ and in the slope of the curves across the
kinematic region.

Among the holographic frameworks considered, the IMA model yields
the largest values of the form factor over the entire range of
$q^{2}$, showing a good level of compatibility with the LCSR
predictions of Ref. \cite{HanWuFu}. In contrast, the 't Hooft model
provides the lowest values throughout, indicating a more suppressed
contribution from nonperturbative dynamics in this description.

The LMZV prediction lies in an intermediate region and shows clear
overlap with the lower cluster of theoretical results, such as QCDSR
\cite{AliAzi} and the CLF model \cite{CheChuHwa}. On the other hand,
some LCSR-based evaluations, including Refs.
\cite{Khosravi1,ZGWang}, as well as PQCD \cite{RunHui} predict
substantially larger form factor values. These discrepancies
originate from the different treatments of long-distance QCD
effects: LCSR calculations are sensitive to the shape of the
light-cone DAs and may enhance endpoint contributions, while
holographic models incorporate confinement and hadronic wave
function structure through entirely different mechanisms.
Consequently, the spread among the various predictions reflects the
intrinsic sensitivity of $f_{+}(q^{2})$ to the modeling of
nonperturbative bound-state effects.

The rare semileptonic decays $B \to K_0^*(1430)\, \ell^+ \ell^- \,
(\ell=\mu,\tau)$ proceed via the flavor-changing neutral current
(FCNC) transition $b \to s$ at the loop level within the Standard
Model (SM). Employing the effective Hamiltonian formalism and
expressing the hadronic matrix elements in terms of the form factors
introduced above, the differential decay width in the rest frame of
the $B$ meson can be written as:
\begin{equation}\label{eq524}
\frac{d\Gamma}{dq^2}\big( B \rightarrow K_0^*\, \ell^+ \ell^-\big) =
\frac{G_{F}^2{\mid}V_{tb}V^*_{ts}{\mid}^2 m_{B}^3\alpha^2}{3\times
2^9\pi ^5}v \,\phi^{1/2}(1,\hat{r},\hat{s})\Big[
\Big(1+\frac{2\hat{l}}{\hat{s}}\Big) \phi(1,\hat{r},\hat{s})\alpha
_1+12\,\hat{l}\beta_1\Big],
\end{equation}
Here, $\hat{r} = m_{K_0^*}^2/m_{B}^2$, $\hat{s}=q^2/m_{B}^2$,
$\hat{m}_b=m_b/m_{B}$, $\hat{l}=m_{l}^2/m_{B}^2$,
$v=\sqrt{1-4\hat{l}/\hat{s}}$, and $\phi(1,\hat{r},\hat{s})=
1+\hat{r}^2+\hat{s}^2-2\hat{r}-2\hat{s}-2\hat{r}\hat{s}$. The
functions $\alpha_1$ and $\beta_1$ encode the short-distance Wilson
coefficients and the hadronic form factors.
\begin{align}
\alpha_1 &= \bigg|C_{9}^{\mathrm{eff}}\,f_{+}(q^2)
+\frac{2\,\hat{m}_b\,
C_{7}^{\mathrm{eff}}\,f_{T}(q^2)}{1+\sqrt{\hat{r}}}\bigg|^{2}
+\big|C_{10}f_{+}(q^2)\big|^{2}, \label{eq525}\\
\beta_1 &= |C_{10}|^{2}\Bigg[ \left(
1+\hat{r}-\frac{\hat{s}}{2}\right) |f_{+}(q^2)|^{2}+\left(
1-\hat{r}\right) \mathrm{Re}[f_{+}(q^2)f^*_{-}(q^2)]
+\frac{1}{2}\hat{s}\,|f_{-}(q^2)|^{2}\Bigg]. \label{eq526}
\end{align}
In the numerical analysis, we retain the short-distance
contributions encoded in $C_9^{\mathrm{eff}}$ and neglect possible
long-distance resonance effects associated with intermediate
charmonium states, restricting the analysis to regions sufficiently
far from the $J/\psi$ and $\psi'$ peaks.

For the numerical evaluation, we use $\tau_{B}=(1.638 \pm
0.004)\,\mathrm{ps}$, $m_\mu = 0.105$ GeV, $m_\tau = 1.776$ GeV
\cite{PDG}, together with the Wilson coefficients
$C_{7}^{\mathrm{eff}}=-0.313$, $C_{10}=-4.669$ \cite{GriSavWise},
and $C_{9}^{\mathrm{eff}}=C_{9}+Y_{S}(q^2)$, where $Y_S(q^2)$
accounts for perturbative contributions of four-quark operators away
from resonance regions \cite{BurMunz}.

\begin{figure}[h]
\begin{center}
\includegraphics[width=8.75cm,height=8cm]{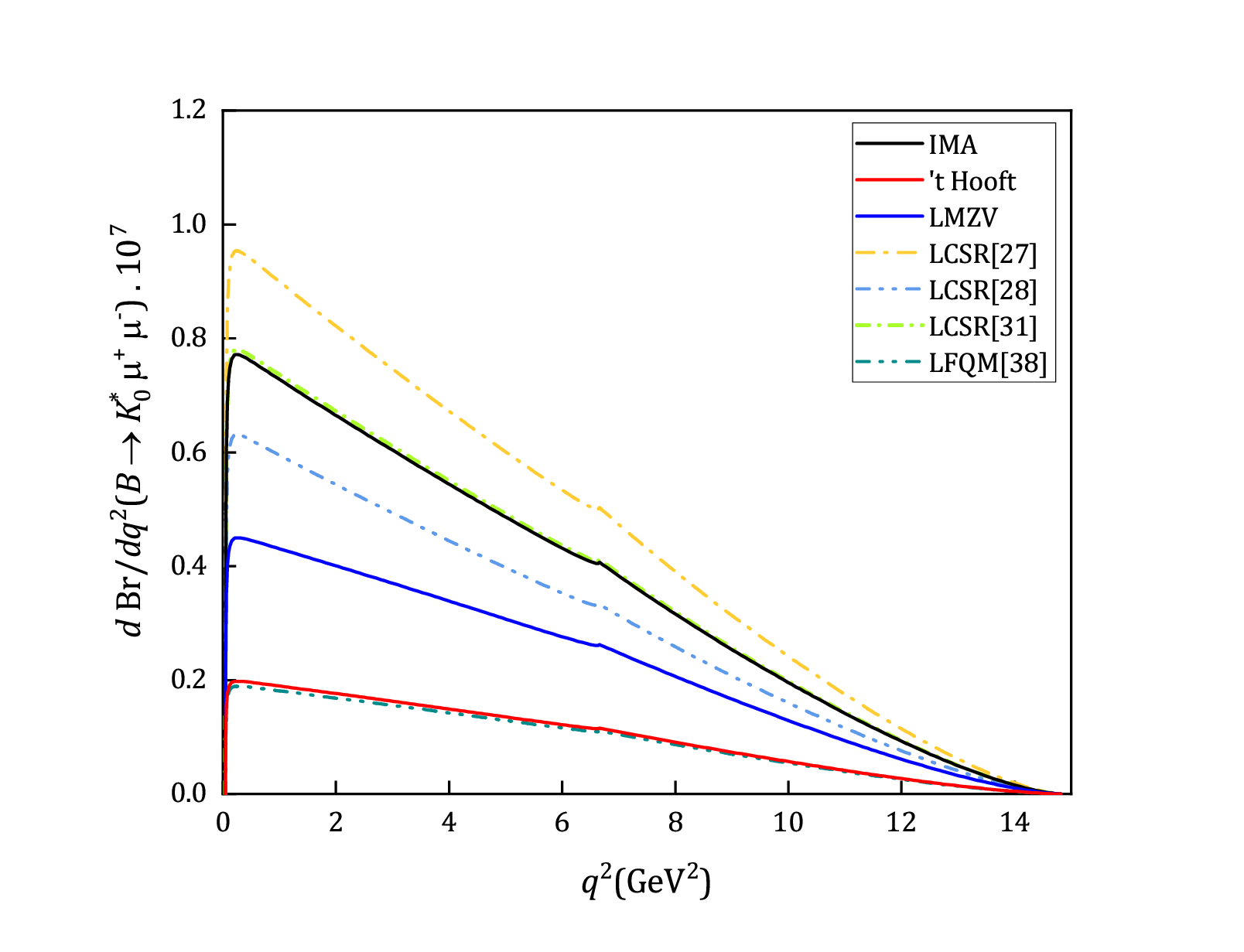}
\includegraphics[width=8.75cm,height=8cm]{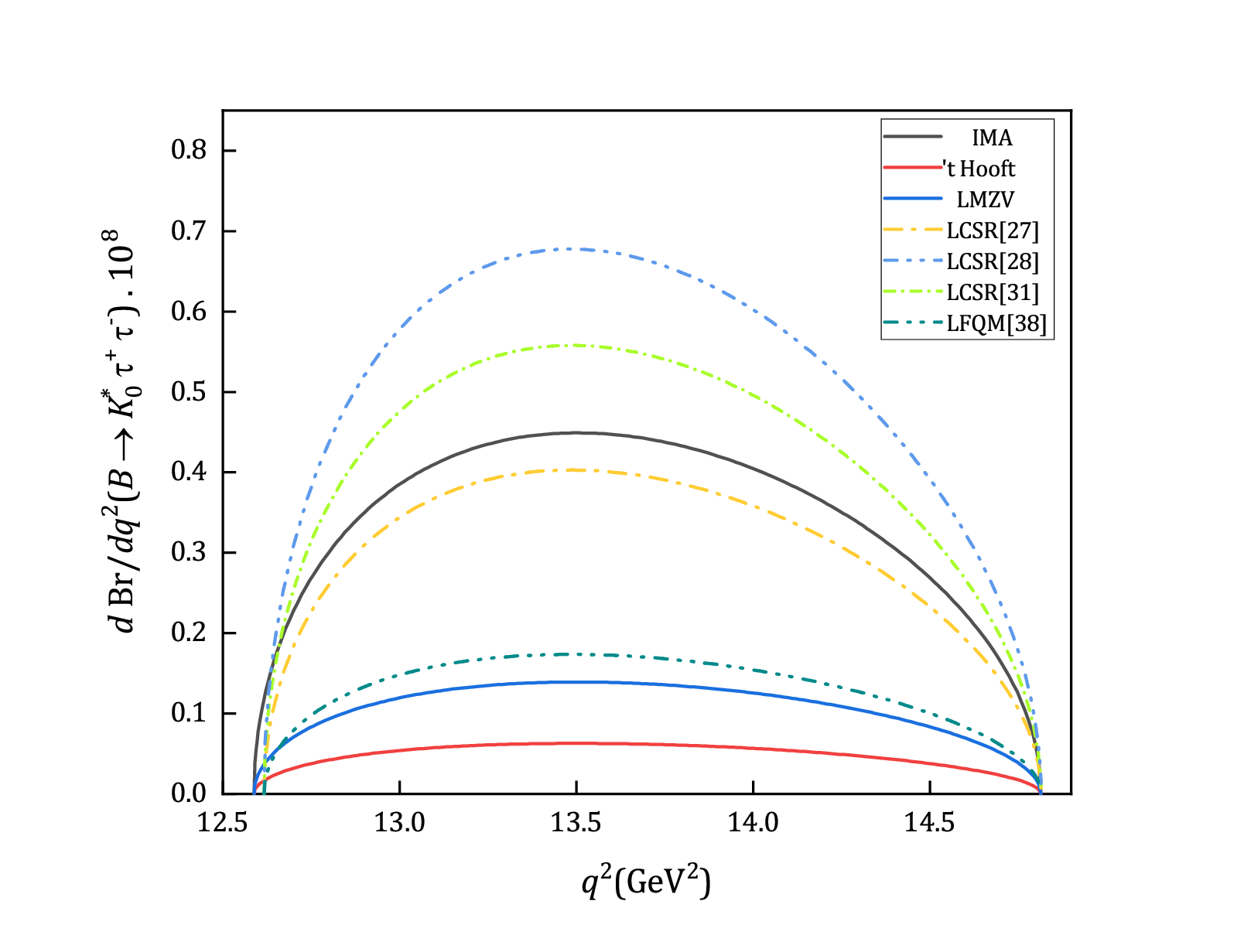}
\caption{The dependence of the differential branching ratios of the
semileptonic $B \to K_0^*\, \ell^+ \ell^-$ decays ($ \ell=\mu,\tau$)
on $q^2$ in different models.} \label{F509}
\end{center}
\end{figure}

Figure \ref{F509} displays the $q^{2}$ dependence of the
differential branching ratios for the semileptonic decays $B \to
K_0^* \ell^+\ell^-$ ($\ell=\mu,\tau$) obtained within our three
holographic frameworks, namely IMA, 't Hooft, and LMZV, and compares
them with several nonperturbative results available in the
literature, including LCSR \cite{Khosravi1,ZGWang,WanAsl}, and LFQM
\cite{ChGeLi}.
As can be seen from the figure, the main phenomenological outcome of
the analysis is not limited to an integrated prediction, but rather
involves the full $q^{2}$ profile of the differential branching
ratio, which provides a more sensitive probe of model dependence
across the physical region.

At the qualitative level, the different approaches exhibit a broadly
similar behavior as functions of $q^{2}$, indicating that they
capture the same general kinematic trend for the decay rate.
Nevertheless, visible differences remain among the various curves,
reflecting the sensitivity of the branching ratio distributions to
the underlying form factors and, more fundamentally, to the
treatment of nonperturbative hadronic dynamics in each framework. In
this respect, the spread among the three holographic predictions
themselves may be viewed as an indicator of the residual model
dependence associated with the longitudinal dynamics implemented in
the holographic description.

The figure also reveals a clear lepton-channel dependence. In
particular, the $\tau^+\tau^-$ mode is confined to a narrower
high-$q^2$ region and appears with a smaller differential rate than
the corresponding $\mu^+\mu^-$ channel. This feature is consistent
with the stronger kinematic suppression expected in the heavy-lepton
case and further illustrates the importance of presenting the
results differentially in $q^{2}$. Overall, the comparison shown in
Fig. \ref{F509} provides a useful benchmark for assessing the
phenomenological viability of the holographic predictions and their
level of agreement with other established approaches.

Integrating Eq. (\ref{eq524}) over the kinematically allowed region
$4m_\ell^2 \le q^2 \le (m_B - m_{K_0^{*}})^2$, we obtain the
branching ratios for the rare decay $B \rightarrow
K_0^{*}\ell^{+}\ell^{-}$. The resulting predictions within our
holographic framework are presented in Table \ref{T508}. For
comparison, the corresponding results from other approaches,
including LCSR \cite{Khosravi1,ZGWang,WanAsl}, QCDSR \cite{AliAzi},
PQCD \cite{RunHui}, and LFQM \cite{ChGeLi}, are also listed.
\begin{table}[ht]
\caption{Branching ratios for the decays $B\to
K_{0}^{*}\,\ell^{+}\ell^{-}$ with $\ell=\mu$ and $\ell=\tau$
obtained in our holographic models, together with predictions from
other approaches.} \label{T508}
\begin{ruledtabular}
\begin{tabular}{lccccccccc}
Observable & IMA & \text{'t Hooft} & LMZV & LCSR \cite{Khosravi1} &
LCSR \cite{ZGWang}& LCSR \cite{WanAsl} &
QCDSR \cite{AliAzi} & PQCD \cite{RunHui} & LFQM \cite{ChGeLi} \\
\hline $\mathrm{Br}(B \to K_{0}^{*}\,\mu^{+}\mu^{-})\times10^{7}$ &
$5.39^{+2.05}_{-1.13}$ & $1.60^{+0.47}_{-0.31}$ &
$3.26^{+0.78}_{-0.57}$ & $5.66^{+2.77}_{-1.98}$ &
$4.12^{+1.35}_{-1.35}$ & $5.60^{+3.10}_{-2.30}$ & $2.07\!-\!2.66$ &
${9.78}^{+7.66}_{-4.40}$
& $1.62$ \\
$\mathrm{Br}(B \to K_{0}^{*}\,\tau^{+}\tau^{-})\times10^{9}$ &
$6.98^{+5.57}_{-2.42}$ & $1.52^{+1.43}_{-0.49}$ &
$2.53^{+2.01}_{-0.67}$ & $5.50^{+3.60}_{-2.80}$ & $11^{+7}_{-7}$ &
$9.80^{+12.40}_{-5.50}$ & $1.7\!-\!2.20$ & ${6.30}^{+5.70}_{-3.00}$
& $2.86$ \\
\end{tabular}
\end{ruledtabular}
\end{table}

For the muonic channel $B \to K_0^{*}\, \mu^{+}\mu^{-}$, the
numerical results display a clear hierarchy among the three
holographic scenarios. The IMA potential yields the largest
branching ratio, while the 't Hooft potential gives the smallest
value, with the LMZV prediction lying in between. Since the
short-distance Wilson coefficients and the input parameters are
identical in all three cases, this pattern originates primarily from
the different $q^{2}$-dependence of the transition form factors
encoded in the corresponding light-front holographic wave functions.

A quantitative comparison with the values collected in Table
\ref{T508} confirms this trend. The IMA result,
$5.39^{+2.05}_{-1.13}\!\times\!10^{-7}$, is in excellent agreement
with the LCSR estimates $(4.12\!-\!5.66)\times10^{-7}$. The 't Hooft
prediction, $1.60^{+0.47}_{-0.31}\!\times\!10^{-7}$, matches well
the QCDSR interval $2.07$--$2.66$ and is also close to the LFQM
value $1.62$, while the LMZV result overlaps with the lower portion
of the LCSR ranges. The PQCD prediction shows a tendency toward
larger central values but comes with sizable theoretical
uncertainties, which softens the apparent deviation.

A similar hierarchy is found in the tauonic channel $B \to
K_0^{*}\,\tau^{+}\tau^{-}$. The IMA prediction,
$6.98^{+5.57}_{-2.42}\!\times\!10^{-9}$, remains compatible with the
broad LCSR ranges, whereas the 't Hooft and LMZV results fall within
the QCDSR interval $1.7$--$2.20$ and are numerically close to the
LFQM value $2.86$. The PQCD approach again yields larger central
values accompanied by substantial uncertainties. Due to the large
$\tau$-lepton mass, the available phase space is significantly
reduced, suppressing the decay rate and enhancing the sensitivity of
the predictions to the details of the form factors; as a result, the
theoretical uncertainties in this channel are typically larger than
in the muonic case.

Overall, the pattern observed in Table \ref{T508} demonstrates that
the spread among our three holographic scenarios is driven mainly by
the different $q^{2}$-dependence of the transition form factors. The
consistent overlap with the LCSR, QCDSR, LFQM, and- within
uncertainties- PQCD predictions underlines the robustness and
phenomenological viability of our results.

The predicted branching ratios indicate that the $B \to
K_0^{*}\,\ell^{+}\ell^{-}$ decays may lie within the experimental
sensitivity of present and forthcoming flavor experiments such as
LHCb and Belle II, providing a useful probe of the hadronic dynamics
governing the $B \to K_0^{*}$ transition.

The longitudinal lepton polarization asymmetry provides a sensitive
probe of the chiral structure of the effective Hamiltonian governing
the $b \to s\,\ell^{+}\ell^{-}$ transition. For the decay $B \to
K_0^{*}\,\ell^{+}\ell^{-}$, the longitudinal polarization asymmetry
$P_L$ is defined as
\begin{equation}\label{eq527}
P_L=\frac{2v}{\Big(1+\frac{2\hat{l}}{\hat{s}}\Big)\phi(1,\hat{r},\hat{s})\alpha_1+12
\hat{l}\beta_1}
\mathrm{Re}\Bigg[\phi(1,\hat{r},\hat{s})\Big(C_9^{\mathrm{eff}}f_{+}(q^2)-\frac{2C_{7}^{\mathrm{eff}}
f_T(q^2)}{1+\sqrt{\hat{r}}} \Big)(C_{10}f_{+}(q^2))^{*}\Bigg],
\end{equation}
Here, the kinematical variables $v$, $\hat{l}$, $\hat{r}$,
$\hat{s}$, the function $\phi(1,\hat{r},\hat{s})$, and the auxiliary
quantities $\alpha_1$ and $\beta_1$ have been defined previously.

\begin{figure}[th]
\begin{center}
\includegraphics[width=8.75cm,height=8cm]{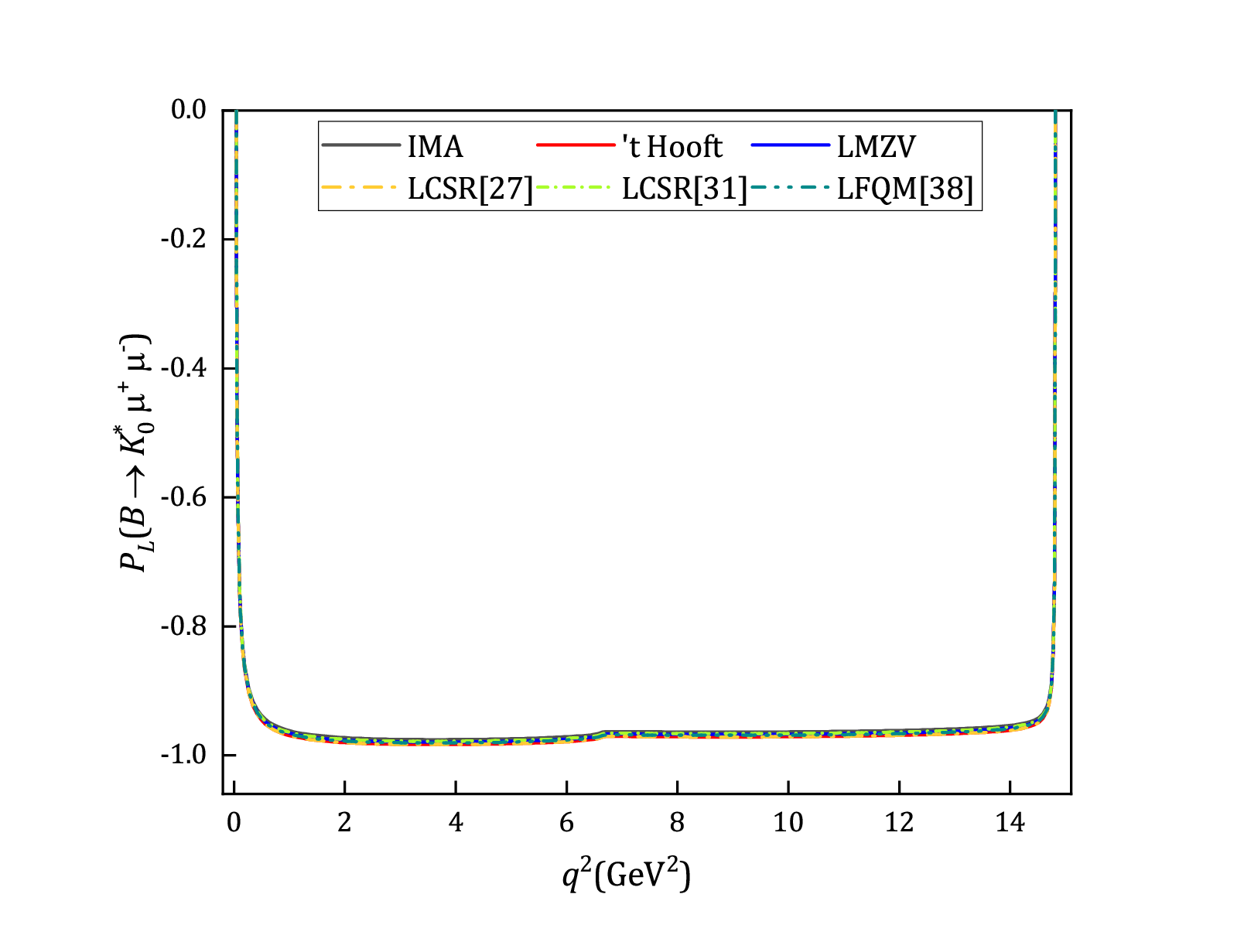}
\includegraphics[width=8.75cm,height=8cm]{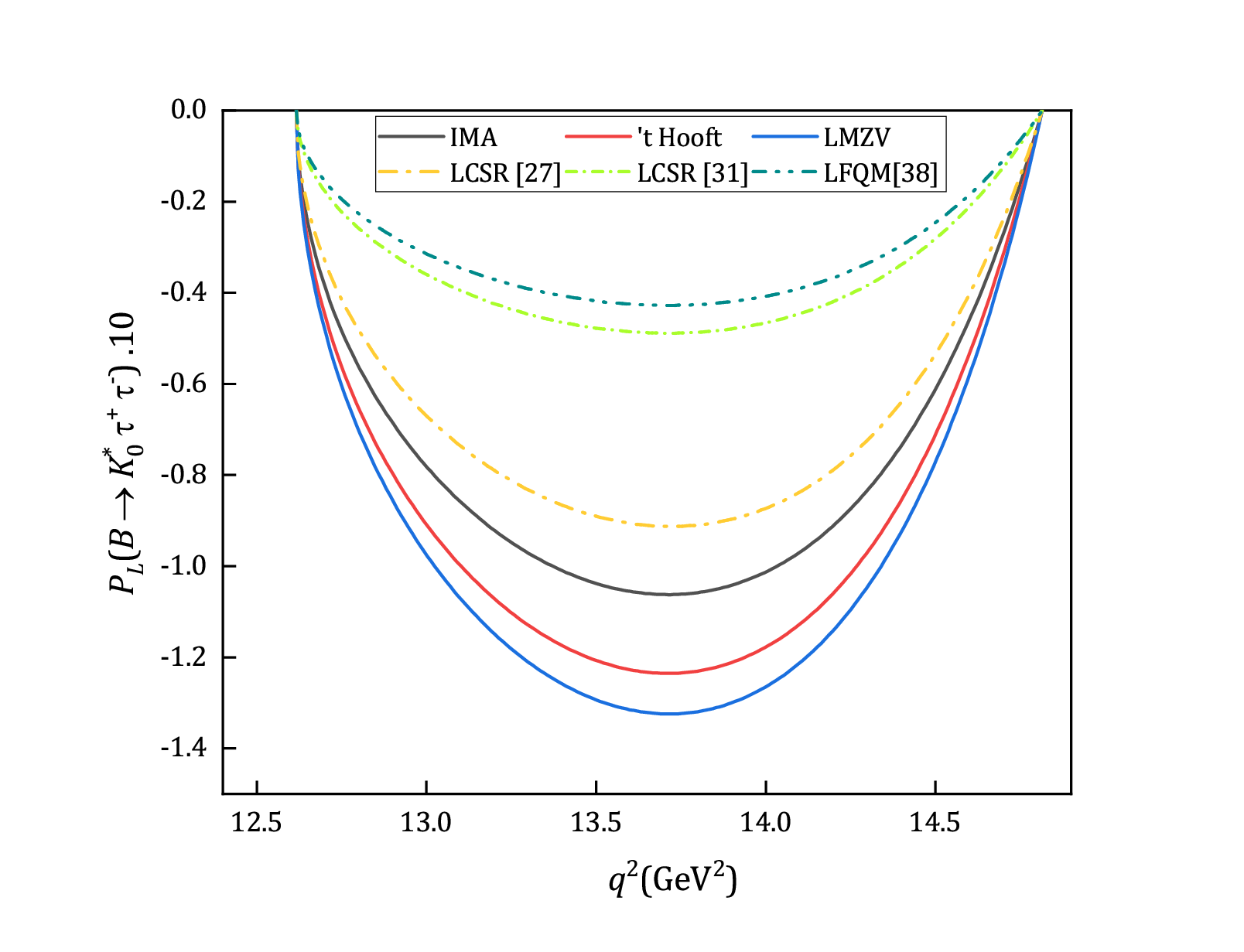}
\caption{The $q^2$ dependence of the longitudinal lepton
polarization asymmetry $P_L$ for the decays $B \to K_0^{*}
\ell^{+}\ell^{-}$ within the IMA, 't Hooft, and LMZV holographic
models. The results are compared with the predictions from other
theoretical approaches.} \label{F510}
\end{center}
\end{figure}

The $q^2$ dependence of the longitudinal polarization asymmetry
$P_L$ for the decays $B \to K_0^{*}\,\ell^{+}\ell^{-}$ with
$\ell=\mu, \tau$ is displayed in Fig.~\ref{F510} for the three
holographic models considered in this work.

In the muonic channel, the longitudinal polarization asymmetry
remains very close to $-1$ across most of the kinematically allowed
region, deviating only near the endpoints of the $q^2$ spectrum.
This behavior reflects the dominance of the left-handed structure of
the $b \to s\,\ell^{+}\ell^{-}$ transition in the SM, primarily
governed by the axial-vector Wilson coefficient $C_{10}$, which
largely determines the helicity structure of the decay amplitude.
Moreover, the $q^2$ dependence of $P_L$ in the muonic mode is nearly
identical for the three holographic scenarios, indicating that this
observable is largely insensitive to moderate variations in the
nonperturbative input.

For the tauonic channel $B \to K_0^{*}\,\tau^{+}\tau^{-}$, the
deviation of $P_L$ from $-1$ becomes more pronounced, especially in
the intermediate-$q^2$ region. This originates from the sizable mass
of the $\tau$ lepton, which partially relaxes the helicity
suppression present in the muonic case and consequently reduces the
magnitude of the longitudinal polarization asymmetry throughout the
entire $q^2$ domain. As a result, this observable becomes more
sensitive to kinematical constraints than to the detailed $q^2$
dependence of the transition form factors.

The averaged longitudinal lepton polarization asymmetries are
presented in Table \ref{T509} for the three holographic light-front
models, along with representative results from the LCSR
\cite{Khosravi1,WanAsl}, and LFQM \cite{ChGeLi} approaches.
\begin{table}[t]
\caption{Averaged longitudinal lepton polarization asymmetries
$\langle P_L \rangle$ for the decays $B \to
K_0^*(1430)\,\ell^{+}\ell^{-}$ ($\ell=\mu,\tau$) obtained within the
three holographic scenarios. The results are compared with the LCSR
\cite{WanAsl} and LFQM \cite{ChGeLi} predictions.} \label{T509}
\begin{ruledtabular}
\begin{tabular}{lccccc}
Channel & IMA & \text{'t Hooft} & LMZV & LCSR \cite{WanAsl} & LFQM \cite{ChGeLi} \\
\hline
$B \to K_0^*\,\mu^{+}\mu^{-}$   & $-0.98$ & $-0.97$ & $-0.97$ & $-0.96$ & $-0.95$ \\
$B \to K_0^*\,\tau^{+}\tau^{-}$ & $-0.09$ & $-0.11$ & $-0.12$ & $-0.03$ & $-0.03$ \\
\end{tabular}
\end{ruledtabular}
\end{table}

For the muonic mode, $\langle P_L \rangle$ stays very close to $-1$
across all approaches, taking values between $-0.95$ and $-0.98$.
This remarkable stability under different theoretical inputs
demonstrates the dominance of the left-handed structure of the
effective Hamiltonian and confirms the limited sensitivity of this
observable to nonperturbative uncertainties.

In contrast, the tauonic channel exhibits a pronounced suppression
of the averaged longitudinal polarization asymmetry, with $|\langle
P_L \rangle| \sim 0.1$ in all considered approaches. While the three
holographic light-front models predict slightly different numerical
values, their results remain mutually consistent and cluster around
a commonly suppressed magnitude. In comparison, the LCSR
\cite{Khosravi1,WanAsl}, and LFQM \cite{ChGeLi} predictions yield
somewhat smaller absolute values of $\langle P_L \rangle$, yet they
clearly confirm the same strong suppression relative to the muonic
mode. This behavior originates from the sizable $\tau$-lepton mass,
which significantly modifies the helicity structure of the decay
amplitude by enhancing the contribution of the lepton-mass-dependent
terms. As a result, the overall longitudinal polarization is
strongly reduced, rendering $\langle P_L \rangle$ in the tauonic
channel more sensitive to kinematical effects than to details of the
nonperturbative hadronic inputs.

\section*{Conclusion}
In this work, we have presented a comprehensive light-front
holographic study of the scalar meson $K_0^*(1430)$ and its role in
the rare semileptonic decay $B \to K_0^*(1430)\,\ell^+\ell^-$. Our
analysis was carried out within three distinct realizations of the
longitudinal mode of the LFWF, namely the IMA, the dynamical 't
Hooft model, and the LMZV approach. These frameworks provide,
respectively, phenomenological, nonlocal dynamical, and analytically
solvable implementations of longitudinal confinement, and therefore
allow for a systematic investigation of how the longitudinal
structure of the meson affects hadronic observables. In this sense,
the present work goes beyond a single-model analysis and establishes
a unified framework for comparing effective and dynamical
descriptions of confinement in the scalar sector.

A first important outcome of our study is that the longitudinally
improved holographic framework provides a quantitatively consistent
description of the kaon family, including the scalar state
$K_0^*(1430)$. In particular, the mass predictions obtained in the
't Hooft and LMZV approaches for $K_0^*(1430)$ are in very good
agreement with the available PDG data, and the corresponding
Regge-like behavior of the kaon trajectories supports the adopted
assignment of quantum numbers. This result strengthens the physical
credibility of the constructed LFWFs and provides a solid foundation
for the subsequent calculation of nonperturbative observables.

Based on these wave functions, we computed a broad and correlated
set of quantities, including the decay constants, the twist-$2$ and
twist-$3$ DAs, their $\xi$-moments, and the valence PDF of the
$K_0^*(1430)$ meson. This combined analysis is one of the central
strengths of the present work, since it connects the internal
light-front structure of the scalar meson to several independent
observables within the same theoretical setup. Our results show
that, although all three approaches reproduce the expected endpoint
behavior and the characteristic asymmetry associated with the
strange-light quark system, the detailed shapes of the longitudinal
wave functions and the resulting DAs remain sensitive to the
underlying implementation of longitudinal confinement. In
particular, the IMA framework generally leads to broader
distributions, whereas the 't Hooft and LMZV models produce more
localized profiles and exhibit a close numerical resemblance,
reflecting their common dynamical origin.

The analysis of the DAs and their $\xi$-moments further clarifies
the pattern of model dependence. We find that the dominant
sensitivity to the longitudinal dynamics is concentrated in the
antisymmetric part of the leading-twist DA, as reflected by the odd
$\xi$-moments, while the even moments and the symmetric core remain
comparatively stable across different models. The twist-$3$
amplitudes also display nontrivial quantitative differences,
especially in their peak structure, indicating that higher-twist
observables provide an additional and sensitive probe of the
longitudinal confinement mechanism. Likewise, the valence PDF
exhibits a clear asymmetry driven by $SU(3)$ flavor-symmetry
breaking, with the strange quark carrying a larger fraction of the
longitudinal momentum, while the width and localization of the
distribution remain strongly model dependent. Taken together, these
results show that the longitudinal sector of the holographic wave
function leaves clear and traceable imprints on the nonperturbative
structure of the scalar meson.

An important phenomenological achievement of this work is the use of
these holographic distribution amplitudes as nonperturbative input
for the light-cone sum-rule calculation of the $B \to K_0^*(1430)$
transition form factors. In this way, the study builds a direct
bridge between the internal structure of the scalar meson and
experimentally relevant flavor observables. The predicted form
factors obtained from the three longitudinal frameworks display a
common qualitative behavior, while preserving quantitative
differences that reflect the model dependence already observed at
the level of the wave functions and DAs.

Using the resulting form factors, we further evaluated the
differential and integrated branching ratios, as well as the
longitudinal lepton-polarization asymmetries, for the decays $B \to
K_0^*(1430)\,\mu^+\mu^-$ and $B \to K_0^*(1430)\,\tau^+\tau^-$. The
corresponding predictions are broadly compatible with the range of
existing theoretical estimates from LCSR, QCD sum rules, PQCD, LFQM,
and related approaches, while at the same time providing a clearer
interpretation of the role played by longitudinal confinement. In
particular, the branching fractions remain within experimentally
relevant ranges, making these decays promising channels for future
measurements at facilities such as LHCb and Belle-II. The
polarization analysis also reveals the expected contrast between the
muon and tau channels: the muonic mode remains close to maximal
longitudinal polarization, while the tau mode is substantially
reduced by kinematical effects, thereby offering complementary
sensitivity to hadronic and leptonic dynamics.

Overall, the present work demonstrates that the longitudinally
improved light-front holographic framework provides a coherent,
flexible, and phenomenologically competitive description of the
scalar meson $K_0^*(1430)$. By combining spectroscopy, light-front
wave functions, decay constants, distribution amplitudes,
$\xi$-moments, valence PDFs, transition form factors, and rare-decay
observables in a single unified analysis, we have shown that the
longitudinal dynamics is not a secondary ingredient, but rather a
structurally important component in the description of scalar
hadrons. Our results indicate that the rare decay $B\rightarrow
K_0^*(1430)\ell^+\ell^-$ provides a theoretically well-motivated
framework for probing the nonperturbative longitudinal structure
encoded in holographic QCD. Although no dedicated experimental
measurements are currently available, the ongoing flavor programs at
LHCb and Belle II make this channel a promising target for future
studies. Consequently, the predicted form factors, differential
branching ratios, and longitudinal lepton polarization asymmetries
presented in this work may serve as useful theoretical benchmarks
for future experimental analyses.

\end{document}